\documentclass[useAMS, usenatbib]{mnras}
\usepackage{times}
\usepackage{graphicx}
\usepackage{color}
\usepackage{epstopdf}
\usepackage{amsmath}
\usepackage{amssymb}
\usepackage{datetime}

\def\mnras{MNRAS}
\def\apj{ApJ}
\def\apjs{ApJS}
\def\apjl{ApJL}
\def\nat{Nature}

\def \msun{\rm M_{\odot}}

\begin{document}

\title[DM Halo Shapes in Illustris vs Illustris-Dark]{Shape of Dark Matter Haloes in the Illustris Simulation: \\Effects of Baryons}

\author[K. E. Chua et al.]{Kunting Eddie Chua$^{1,2}$\thanks{Email: kchua@cfa.harvard.edu}, Annalisa Pillepich$^1$$^,$$^3$, Mark Vogelsberger$^4$, and Lars Hernquist$^1$\\
$^{1}$Harvard-Smithsonian Center for Astrophysics, 60 Garden Street, Cambridge, MA 02138\\
$^{2}$Institute of High Performance Computing, 1 Fusionopolis Way, Singapore 138632\\
$^{3}$Max-Planck-Institut f{\"u}r Astronomie, K{\"o}nigstuhl 17, 69117 Heidelberg, Germany\\
$^{4}$Massachusetts Institute of Technology, Cambridge, MA 02138}

\maketitle
\begin{abstract}
We study the effect of baryonic processes on the shapes of dark matter (DM) haloes from Illustris, a suite of hydrodynamical (Illustris) and DM-only (Illustris-Dark) cosmological simulations performed with the moving-mesh code {\sc arepo}. 
DM halo shapes are determined using an iterative method based on the inertia tensor for a wide range of $z=0$ masses ($M_{200} = 1 \times 10^{11} - 3 \times 10^{14} M_\odot$). 
Convergence tests shows that the local DM shape profiles are converged only for $r > 9\epsilon$, $\epsilon$ being the Plummer-equivalent softening length, larger than expected. 
Haloes from non-radiative simulations (i.e. neglecting radiative processes, star formation, and feedback) exhibit no alteration in shapes from their DM-only counterparts: 
thus moving-mesh hydrodynamics alone is insufficient to cause differences in DM shapes. 
With the full galaxy-physics implementation, condensation of baryons results in significantly rounder and more oblate haloes, with the median minor-to-major axis ratio $\left< s\equiv c/a \right> \approx 0.7$, almost invariant throughout the halo and across halo masses. 
This somewhat improves the agreement between simulation predictions and observational estimates of the Milky Way halo shape. 
Consistently, the velocity anisotropy of DM is also reduced in Illustris, across halo masses and radii.
Within the inner halo ($r=0.15 R_{200}$),  both $s$ and $q$ (intermediate-to-major axis ratio) exhibit non-monotonicity with galaxy mass, peaking at $m_* \approx 10^{10.5-11} M_\odot$,
which we find is due to the strong dependence of inner halo shape with galaxy formation efficiency. 
Baryons in Illustris affect the correlation of halo shape with halo properties, leading to a positive correlation of sphericity of MW-mass haloes with halo formation time and concentration, the latter being mildly more pronounced than in Illustris-Dark. 
\\
\\
\end{abstract}

\section{Introduction}
Under the hierarchical cold dark matter ($\Lambda$CDM) theory of structure formation, large haloes form from the accretion of diffuse matter and by merging with other haloes. Halo growth is generally anisotropic since accretion can be clumpy and directional (e.g. along filaments and sheets), resulting in the formation of non-spherical triaxial haloes. 

Although baryons are an integral part of galaxy formation, due to the difficulty in their modelling, most predictions about the shapes of DM haloes come from numerical $N$-body, dark-matter only (DMO) simulations \citep[e.g.][]{Dubinski91v378, Warren92v399, Dubinski94v431,Jing02v574,Bailin05v627,Allgood06v367,Maccio08v391,Jeeson11v415}, that neglect baryonic processes. These studies showed that CDM haloes are both triaxial and prolate ($c/b > b/a$)\footnote{$a>b>c$ are the major, intermediate and minor axes lengths, respectively, throughout this paper.}. More massive haloes also tend to be slightly less spherical than lower-mass haloes, while more concentrated ones are more spherical. In particular, past and recent studies of Milky Way (MW)-sized haloes in $N$-body simulations predict an average value of the minor-to-major axis ratio $\left< c/a \right> \lesssim 0.5$ within few tens of kpc from the galactic center.

In the Milky Way, work has been done to model the potential and shape of the MW halo  using stellar streams, which can be assumed to trace the MW potential \citep[e.g.][]{Ibata01v551,Law10v714,Vera-Ciro13v773,Bovy16v833}. For example, using the tidal stream of the Sagittarius dwarf spheroidal galaxy, \cite{Ibata01v551} arrived at a value of $\left< c/a \right> \ge 0.8$ while \cite{Law10v714} obtained  $\left< c/a \right> = 0.72$ and $\left< b/a \right> = 0.99$. The incompatibility between these results and those of $N$-body simulations suggest that the MW inner halo (between 16-60 kpc) is likely to be more spherical than $N$-body simulations have predicted.

$N$-body simulations are unable to provide a complete picture of galaxy formation, because the coupling of baryons and DM can have a significant impact on the structure of DM haloes especially in the inner halo where galaxies reside. For example, the condensation of baryons at halo centres can modify the potential wells of haloes, leading to effects such as adiabatic contraction in the central regions where the DM halo concentrations in hydrodynamic simulations are enhanced relative to their $N$-body counterparts \citep{Blumenthal86v301,Gnedin04v616}. On the other hand, stellar and active galactic nucleus (AGN) feedback can expel both baryons and DM from the core, reducing central concentrations instead \citep[e.g.][]{Duffy10v405}. 

In contrast to dissipationless $N$-body simulations, work by \cite{Katz91v377} and \cite{Katz93v412} were first to note the sphericalisation of DM haloes in dissipational simulations. This was followed by \cite{Dubinski94v431} who studied the effects of baryon dissipation on halo shapes by adiabatically growing a galaxy at the centre of initially triaxial DM halo, reaching similar conclusions. Such a sphericalisation can be due to the modification of the orbital structure of a halo, with box orbits that pass close to the centre being scattered by the central galaxy.  These initial works were, however, plagued by low resolutions and by environments that were not representative of the cosmological framework in which haloes actually form and grow.

Further progress has been made in this regard, with new cosmological hydrodynamic simulations being used to analyse DM shapes. These include work by e.g. \cite{Debattista08v681,Tissera10v406,Abadi10v407,Kazantzidis10v720,Bryan13v429,Butsky16v462}.
In particular, \cite{Bryan13v429} studied how halo and galaxy properties affect halo shapes using the OWLs simulations \citep{Schaye10v402}, which is one of the first suites of cosmological simulations aimed at producing realistic galaxy populations. Since OWLs consisted of hydrodynamic simulations with identical initial conditions but with varying stellar and AGN feedback models, \citealt{Bryan13v429} were able to ascertain that changing feedback in the simulations can lead to substantial changes in halo shapes through its effect on the galaxy formation efficiency. Like \cite{Abadi10v407}, they found that baryons make the DM halo more spherical, but that strong stellar and AGN feedback can reduce the impact of baryons. However, the halo shapes in \cite{Bryan13v429} were calculated using the non-iterative method suggested in \cite{Bailin05v627}, which is less accurate than iterative methods \citep{Zemp11v197}, and also does not take into account variations in halo shapes with distance from the halo centre.

In this work, we further investigate and quantify the effect of baryonic physics on DM halo shapes and the relation to halo and galaxy properties by using yet another galaxy-physics model. We compare a hydrodynamical simulation (Illustris) with the $N$-body (DMO) counterpart simulation of identical volume (Illustris-Dark) that are both part of the Illustris project (www.illustris-project.org).
Our hydrodynamical simulation includes processes such as radiative heating and cooling, star formation, chemical evolution as well as strong supernova and AGN feedback.
In \cite{Chua2017}, we found that baryons led to a drastically different concentration--mass relation in Illustris not only when compared to the $N$-body case, but also when compared with other recent hydrodynamic simulations such as EAGLE \citep{Schaller15v451} or IllustrisTNG \citep{Lovell2018}. In light of these differences, it is useful to study the shapes of DM haloes in Illustris, and understand how the halo shapes can reflect the underlying different baryonic physics implementations.  

The paper is organized as follows: 
we describe our simulations and methods in Section 2 and discuss the convergence of shape profiles in Section 3.
We present our results on the effect of baryons on the halo shape in Section 4, with comparisons to observations of the Milky Way in Section 5.
We also examine how halo and galaxy properties drive halo shapes in Section 5, and finally summarise our results in Section 6.

\section{Methods and Definitions}
\subsection{The Illustris Simulations}

In this work, we analyse haloes drawn from from the Illustris project \citep{Illustris,Vogelsberger14v444,Genel14v445,Sijacki15v452}, which consists of a series of 
cosmological simulations with a box-size of 106.5 Mpc a side. The cosmological parameters used are  consistent with the 9-year Wilkinson Microwave Anisotropy Probe (WMAP-9) results, given by $\Omega_m = 0.27$, $\Omega_\Lambda = 0.73$, $\Omega_b = 0.0456$, $\sigma_8 = 0.81$, $n_s = 0.963$ and $h = 0.704$ \citep{Hinshaw2013}.

The full physics (FP) runs of the Illustris suite include hydrodynamics and key physical processes for galaxy formation, and were performed at three different resolutions: $2\times 1820^3$, $2\times 910^3$ and $2\times 455^3$, with an equivalent number of DM and gas elements at the initial conditions.
For comparison to the hydrodynamic runs, we also investigate haloes from a similar set of DM-only (DMO) simulations performed with the same initial conditions and resolutions.  
In addition, non-radiative (NR) simulations with $2\times 910^3$ and $2\times 455^3$ elements were  also performed. Similar to the FP runs, the NR runs include both DM and baryons, but no radiative cooling, star formation and feedback. The important parameters of these simulations are summarized in Table \ref{table:parameters}.

	\begin{table*}
		\begin{tabular*}{\textwidth}{@{\extracolsep{\fill}}l l c  c c c c c}
			\hline
			Name	& Simulation Type & Volume	& DM particles \& cells	& $\epsilon$	& $m_{\rm DM}$		& $m_{\rm baryon}$ \\ 
			& & [${\rm Mpc}^3$]	&					& [kpc]	&  [$10^6 M_\odot$]	& [$10^6 M_\odot$]	\\
			\hline
			Illustris		& Full physics (FP)&106.5		& $2\times 1820^3$& 1.42/0.71 	& $6.26 $	&$1.26 $\\ 
			Illustris-2		& Full physics (FP)&106.5		& $2\times 910^3$	& 2.84/1.42 	& $50.1 $	&$10.1 $\\ 
			Illustris-3 		& Full physics (FP)&106.5		& $2\times 455^3$	& 5.68/2.84 	& $400.8 $	&$80.5 $\\ \hline
			
			Illustris-NR-2		& Non-radiative hydro (NR) &106.5		& $2\times 910^3$	& 2.84/1.42 	& $50.1 $	&$10.1 $\\ 
			Illustris-NR-3 		& Non-radiative hydro (NR) &106.5		& $2\times 455^3$	& 5.68/2.84 	& $400.8 $	&$80.5 $\\ \hline
			
			Illustris-Dark		& Dark-matter only (DMO) &106.5		& $1820^3$		& 1.42/-	& $7.52 $	& - \\ 
			Illustris-Dark-2	& Dark-matter only (DMO) &106.5		& $910^3$		& 2.84/-	& $60.2 $	& - \\ 
			Illustris-Dark-3	& Dark-matter only (DMO) & 106.5		& $455^3$		& 5.68/-	& $481.3 $	& - \\ \hline
		\end{tabular*}
		\caption{Summary of the Illustris simulation runs and the parameters used: 
			(1) simulation name; 
			(2) simulation type;
			(3) volume of simulation box;
			(4) number of cells and particles in the simulation;
			(5) gravitational softening length ;
			(6) mass per DM particle;
			(7) target mass of baryonic cells.
			The first value of the Plummer-equivalent softening length is given for the DM particles which uses a fixed comoving softening length.
			The gas cells use instead an adaptive softening length with floor specified by the second value of $\epsilon$ \citep{Vogelsberger14v444}. 
		}
		\label{table:parameters}
	\end{table*}

The simulations of the Illustris suite were carried out using the \textsc{arepo} code  \citep{Springel09v401}, where the hydrodynamical equations are solved on a moving Voronoi mesh using a finite volume method. This approach is quasi-Lagrangian since the mesh generating points are advected with the local velocity of the fluid, and combines the advantages of both Eulerian and Lagrangian methods \citep{Vogelsberger12v425,Sijacki12v424}. The gravitational forces are computed using a Tree-PM method where long-range forces are calculated on a particle mesh and the short-range forces are calculated using a hierarchical multipole expansion scheme.

In Illustris, the baryonic processes are treated using sub-resolution models, described fully in \cite{Vogelsberger13} and \cite{Torrey14v438}. In summary, we model star formation following \cite{Springel03v339} where the star-forming interstellar medium is described using an effective equation of state and stars form stochastically above a threshold gas density $\rho_{\rm sfr} = 0.13\,{\rm cm^{-3}}$ with timescale $t_{\rm sfr} = 2.2{\rm\,Gyr}$.
In addition, we account for stellar winds which are modelled as kinetic outflows, and AGN feedback, which is required to quench star formation in massive galaxies.  The AGN feedback mechanism includes not only quasar-mode and radio-mode feedback, where the central black hole accretion rate controls energy release into the surrounding gas, but also non-thermal and non-mechanical electromagnetic feedback. The subgrid parameters have been chosen to reproduce observables such as the cosmic star-formation rate density, galaxy stellar mass function, and the stellar mass - halo mass relation of galaxies. With the galaxy formation implementation, Illustris has been able to achieve good agreements with a broad number of observations at low redshift and across cosmic time \citep[][and results at http://www.illustris-project.org/results/]{Vogelsberger14v444,Genel14v445}.

At each of the 136 simulation snapshots, haloes are identified using a friends-of-friends ({\sc fof}) group finder \citep{Davis85} with a linking length of 0.2. Gravitationally self-bound subhaloes and are subsequently identified using the \textsc{subfind} algorithm \citep{Springel01v328,Dolag09v399}.
The most massive subhaloes in each {\sc fof} group are classified as {\it centrals} with the remaining subhaloes known as {\it satellites}.
For each halo, we denote $R_{200}$ and $M_{200}$ as the virial radius and virial mass respectively\footnote{$R_{\Delta}$ is the radius within which the enclosed mass density is $\Delta$ times the critical value $\rho_c$ i.e. $\rho_{\rm halo} = \Delta \rho_c$. $M_{\Delta}$ is the total mass of the halo enclosed within $R_{\Delta}$ where we choose $\Delta = 200$.}.

\subsection{Halo Matching}

To facilitate comparison between Illustris and Illustris-Dark, we match the (sub)haloes between the two simulations using the unique IDs of the DM particles.
The precise strategy is described in detail in \cite{RodriguezGomez2016} and is based solely on the {\sc subfind} catalogue. 
For any given halo in Illustris, the matching (sub)halo in Illustris-Dark is the (sub)halo that contains the largest fraction of these IDs.
The process can be repeated starting from a (sub)halo in Illustris-Dark to find a match in Illustris.
The final matched catalogue consists of only (sub)haloes with successful matches in both directions.

\subsection{Halo Shape}
\label{sec:iterative}

Since DM haloes are triaxial, their shapes can be described by the axis ratios $q\equiv b/a$ and $s\equiv c/a$ 
where $a$, $b$ and $c$ are the major, intermediate and minor axes respectively
\citep[e.g.][]{Bailin05v627,Allgood06v367}.
The ratio of the minor-to-major axis $s$, has traditionally been used 
as the canonical measure of halo sphericity.

An important quantity required in computing the parameters $q$ and $s$ is the shape tensor $S_{ij}$. Following halo shape literature \citep[e.g.][]{Bailin05v627,Zemp11v197}, we define the {\it shape tensor} as the second moment of the mass distribution divided by the total mass:
\begin{equation}
	S_{ij} = \frac{1}{\sum_k m_k} \sum_k  \frac{1}{w_k} m_k\, r_{k,i} \,r_{k,j}
	\label{eqn:shapetensor}
\end{equation}
where $m_k$ is the mass of the $k$th particle, and $r_{k,i}$ is the $i$th component of its position vector.
$w_k$ is a parameter that can be used to weight the contribution of each particle to $S_{ij}$.
The choice of $w_k$ can be dependent on the aspect of halo shape that is under examination.
Common choices of $w_k$ are $w_k$ = 1 and $w_k = r^2_{\rm ell,k}$ where
\begin{equation}
	r_{\rm ell}^2 = x^2 + \frac{y^2}{(b/a)^2} + \frac{z^2}{(c/a)^2}.
	\label{eqn:rell}
\end{equation}
with $(x,y,z)$ being the position of the particle in its principal frame
and $a$, $b$ and $c$ the lengths of the semi-axes.
For $w_k = 1$, all particles are unweighted and $S_{ij}$ is proportional to the inertia tensor.
For $w_k = r^2_{\rm ell}$, $S_{ij}$ is also known as the {\it reduced inertia tensor} \ and $w_k$ is chosen to reduce the contributions from particles at large distances.

For DM particles, which have fixed mass in the simulations, the shape tensor reduces to 
\begin{equation}
	S_{ij} = \sum_k  \frac{1}{w_k} r_{k,i} \,r_{k,j}.
\end{equation}
This does not hold for baryonic elements that do not have fixed masses.
For stellar shapes, the full shape tensor defined in Equation \ref{eqn:shapetensor} has to be used. In this paper, however, we focus exclusively on the shapes of the DM distribution.

In general, we calculate $q(r)$ and $s(r)$ as a function of distance from the halo centre.
Hence, we fix $w_k = 1 $ and select particles in logarithmic ellipsoidal shells at different distances $r_{\rm ell}$. 
From the equation of an ellipsoidal shell ($1 = x^2/a^2 + y^2/b^2 + z^2/c^2$), 
it is easy to see from Equation \ref{eqn:rell} that $r_{\rm ell}$ is basically its  semi-major length $a$.
In this convention, the ellipsoids and hence the potential of the halo are oriented with $x$ along the longest or {\it major} axis and $z$ along the shortest or {\it minor} axis.

To calculate the shape, the shape tensor is diagonalized to compute its eigenvectors and eigenvalues $\lambda_a$, $\lambda_b$ and $\lambda_c$, with $\lambda_a>\lambda_b>\lambda_c$. The eigenvectors denote the  directions of the principal axes while the eigenvalues are related to the square-roots of the principal axes lengths ($a\propto\sqrt{\lambda_a}$, $b\propto\sqrt{\lambda_b}$ and $c\propto\sqrt{\lambda_c}$; we adopt $a > b > c$, throughout).

Since the shape is unknown a priori, we use an iterative method starting with particles selected in a spherical shell (i.e. $q=s=1$). In each iteration, we select particles in radial bins of width 0.1 dex, diagonalise the shape tensor, and  rotate all particles into the computed principal frame. The process is repeated keeping the semi-major length constant (fixed $r_{\rm ell}$) until $q$ and $s$ converge. For this work, we have chosen a convergence criterion where $q$ and $s$ in successive iteration steps differ by less than 1 per cent.

Since we are in general interested in the shape of the smooth potential of the halo, we avoid substructure contamination by using only particles identified by {\sc subfind} as part of the central subhalo. As such, we neglect substructure and prevent them from biasing the shape calculation.
A discussion of the effects of including substructure in the shape calculation can be found in the Appendix.

There are also instances where it is not the local halo shape at a particular distance but an overall quantification of the shape that is desired. In this case, the shape is calculated for an enclosed volume with the weights $w_k = r^2_{\rm ell}$, using all particles interior to the ellipsoidal surface. Such a procedure biases the shape measurement to interior particles and smooths out shape changes \citep[see][]{Zemp11v197}. We further discuss and show the difference between ellipsoidal shells and volumes in the Appendix.

Another common method for calculating halo shapes involves enclosing a spherical volume 
and diagonalizing the shape tensor without iteration \cite[e.g.][]{Bryan13v429,Bailin05v627}. Such methods require an empirical modification of the axis ratios because it returns values biased towards larger axis ratios due to the use of a spherical volume. Although such procedures are quick to perform, these empirical modifications require careful calibration which can obscure important trends in the results. We avoid such ambiguities by using the iterative procedure and by neglecting substructures, which is most reliable at reproducing local shapes of haloes i.e. when radial profiles are required. For a thorough discussion and comparison of different methods involving the shape tensor, see \cite{Zemp11v197}. 

Finally, the triaxiality parameter, defined as $T\equiv(1-q^2)/(1-s^2)$,
measures the  prolateness or oblateness of a halo. $T=1$ describes a completely prolate halo ($a >  b \approx c$), while $T=0$ describes a completely oblate halo ($a \approx b > c$). In practice, haloes with $T>0.67$ are considered prolate and haloes with $T<0.33$ are oblate. Haloes with $0.33<T<0.67$ are considered triaxial. We refer to the axis ratios $q$ and $s$, and the triaxiality $T$, collectively, as the halo shape parameters.

\subsection{Halo and Galaxy Properties}

Apart from halo mass and shape, other halo properties we consider in this work include:

\begin{itemize}
	
	\item {\it Halo Formation Redshift}, $z_{1/2}$: The halo formation redshift denotes the redshift when a halo has accreted half of its mass at $z=0$ . In practice, we measure $z_{1/2}$  as the earliest moment at which the splined total mass accretion history of a halo reaches half of its $z=0$ mass \citep{Bray16v455} using the halo merger trees derived from the {\sc sublink} merger tree code \citep{Vicente2015v449} . \\
	
	\item {\it Halo Concentration}, $c_{-2}$: We define the halo concentration parameter as $c_{-2}\equiv R_{200}/r_{-2}$. Here, $r_{-2}$ is the scale radius where the slope of the DM density profile takes on the isothermal value i.e. $d \ln \rho/d\ln r =-2$. We obtain $r_{-2}$ by fitting the spherically averaged DM density profile of the halo ($\rho_{\rm DM}(r)$) to an Einasto profile \citep{Einasto}:
	\begin{equation}
		\rho_{\rm DM}(r) = \rho_{-2} \exp\left\{ -2n \left[ \left( \frac{r}{r_{-2}} \right)^{1/n} -1 \right]\right\} 
	\end{equation}
	where $\rho_{-2}$, $n$ are additional fitting parameters. This definition for the concentration differs from the conventional one based on the scale radius of Navarro-Frenk-White profile \citep{Navarro96v462}, and has been found to provide a better description of halo density profiles in hydrodynamical simulations \cite[see e.g.][]{Pedrosa09v395}.\\
	
	\item {\it Halo Velocity Anisotropy}, $\beta(r)$:
	The velocity anisotropy parameter $\beta$ is a measure of anisotropy in the velocity distribution of a halo and can be defined as:
	\begin{equation}
		\beta(r) = 1 - \frac{\sigma_{\rm t}^2 (r)}{2\sigma_{\rm r}^2 (r)}
	\end{equation}
	$\sigma_{\rm r}^2(r) = \left< \left(v_r - \left<v_r \right> \right)^2\right>$  is the (squared) radial velocity dispersion of DM particles in a spherical shell of radius $r$, where $\left<v_r \right>$ is the mean radial velocity in the shell. The tangential velocity dispersion $\sigma_{\rm t}^2$ is defined similarly using the tangential velocity $v_t$. 
	
	A velocity anisotropy of $\beta=0$ corresponds to an isotropic velocity distribution. $\beta > 0$  when radial orbits dominate while $\beta < 0$ when circular orbits dominate. As such, the velocity anisotropy parameter is a useful way to describe the orbital structure of a halo.
	
\end{itemize}

\begin{figure*}
	\centering
	\includegraphics[width=0.95\textwidth]{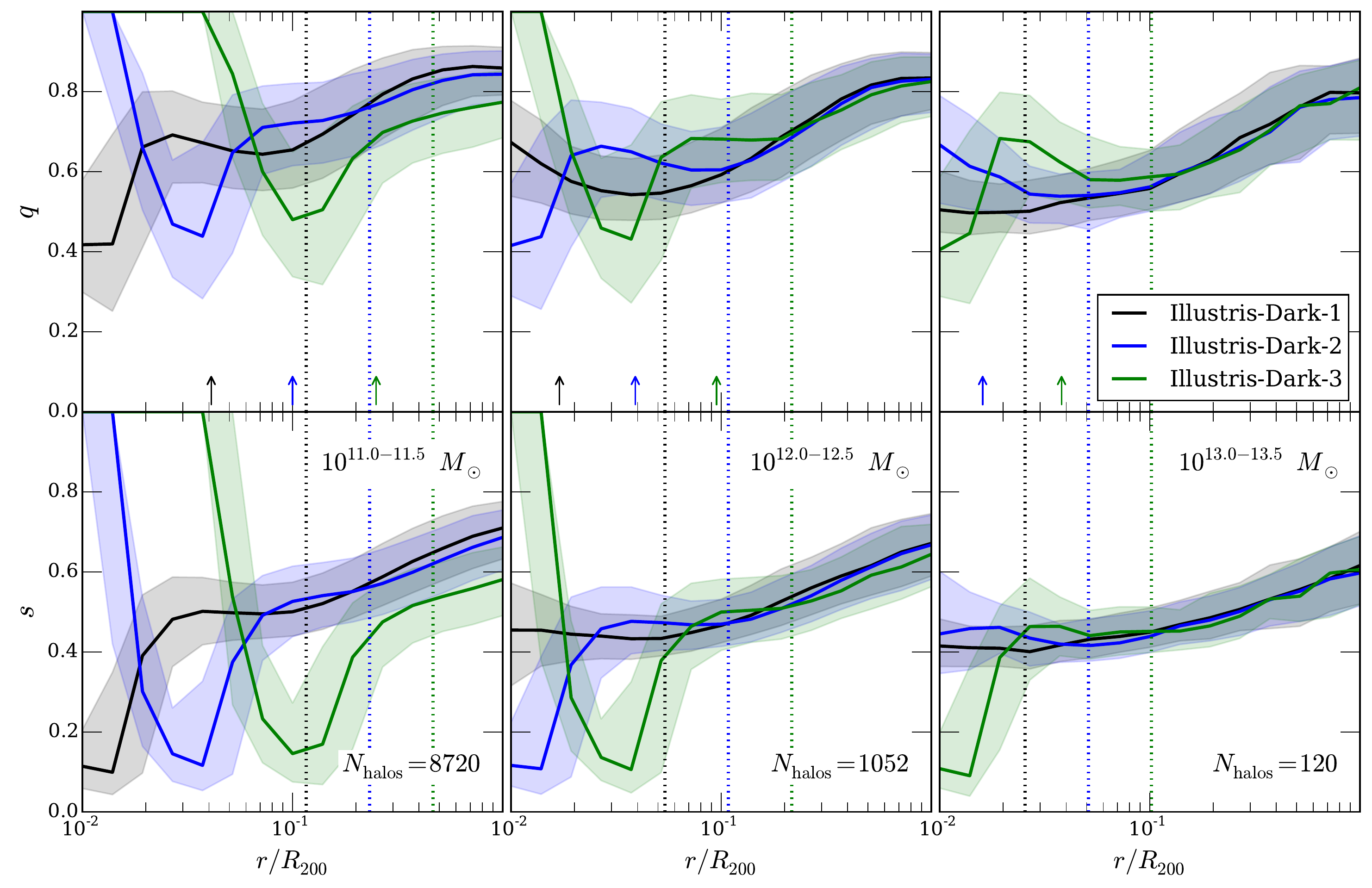}
	\caption{
		Convergence of shape profiles with resolution: Plot of shape parameters $q \equiv b/a$ (top) and $s \equiv c/a$ (bottom) as a function of halocentric distance in the highest resolution DMO run Illustris-Dark-1 (black) and the lower resolution runs Illustris-Dark-2 (blue) and Illustris-Dark-3 (green). Solid lines show the median values while shaded regions show the 25th to 75th central quartile of the galaxy population. The left, middle and right columns correspond to halo masses of $10^{11-11.5}$, $10^{12-12.5}$ and $10^{13-13.5} M_\odot$ respectively. The number of haloes identified in Illustris-Dark-1 is also shown in the bottom row. 	The dotted vertical lines show $9\epsilon$ i.e. 9 times the softening lengths for each resolution, which we consider to be the minimum radii in order to achieve convergence in the shape profiles. For comparison, the convergence criteria of Power et al. (2003) is denoted using coloured arrows. Illustris-Dark shape parameters are well converged down to about 6 per cent of the virial radii.
    }
	\label{fig:resolutiontest}
\end{figure*}

\section{Resolution and Convergence}
\label{sec:resolution}

It is important to understand what regions in a DM halo can be reliably resolved in numerical simulations. The lack of an analytic theory of DM halo structures necessitates the use of convergence studies, as have been applied to  halo mass profiles. For example, \cite{Power2003v338} (hereafter P03) found that the convergence of mass profiles depends on the number of enclosed particles. For convergence, there must be enough particles for the two-body relaxation time-scale to be comparable to the age of the universe.
In most simulations, the halo mass density profiles are converged at $r \gtrsim 3\epsilon$, where $\epsilon$ is the Plummer-equivalent softening length of the DM particles, these criteria applying exclusively to N-body only, DMO, simulations. By considering the P03 criteria, we find this to be approximately true in all three resolutions of Illustris-Dark, with $\epsilon$ shown in Table \ref{table:parameters}.

To understand the convergence of the local shape profiles, we use the three resolution runs of the Illustris suite. Here, we rely on Illustris-Dark for two reasons: 1) to isolate the resolution convergence of the iterative shape procedure described in Section \ref{sec:iterative} and 2) to neglect the resolution effects that are due to baryonic physics in the FP runs. As such, we are not examining here how baryonic physics is affected by resolution.

Figure~\ref{fig:resolutiontest} shows the median shape parameters $q$ (upper panels) and $s$ (lower panels) as a function of halocentric distance for three different halo mass ranges. Colours correspond to different resolutions, with black, blue and green for the highest, medium and lowest resolution runs respectively.
With the exception of the smallest ($10^{11} M_\odot$) haloes in Illustris-Dark-3, the shape profiles of the two lower resolution runs converge with that of Illustris-Dark-1 above some minimum radii.
In general, we find that $q$ and $s$ are converged for $r > 9 \epsilon$, which corresponds to 13, 26 and 51 kpc in Illustris-Dark-1, 2 and 3 respectively.
These convergence radii are shown in  Figure~\ref{fig:resolutiontest} as vertical lines.
While $s(r)$ converges to smaller radii than $q(r)$, we have chosen our resolution criterion to be the more stringent of the two, i.e. using $q(r)$.
For comparison, we have also shown the convergence radii derived from the P03 criterion as arrows in the upper panels of Figure~\ref{fig:resolutiontest}.
We find that minimum converged radii for shapes is between two to three times that of the P03 criterion.
The difference between the convergence of halo shape and spherically average mass profiles is likely a result of the three-dimensional nature of halo shapes compared to the one-dimensional mass profiles. We also varied the width of the ellipsoidal shells between 0.5 dex and 0.25 dex but did not find the width to appreciably affect the obtained median shape profiles, nor their convergence.

For haloes of $10^{11} M_\odot$, we are unable to produced converged shape profiles in the lowest resolution Illustris-Dark-3. In this case, the predicted minimum convergence radii (51 kpc) lies at about 50 per cent of the virial radius. At this resolution, these haloes contain only a few hundred particles within the virial radius, which is insufficient for the halo shape to be resolved.

Convergence studies of halo shapes have been performed in previous work \citep[e.g.][]{Tenneti14v441} and have typically found that at least $\sim$1000 particles is required for the shape calculation to be reliable. However, the difference in procedures between this work and previous studies -- e.g. the use of a unweighted vs. a reduced inertia tensor or the use of ellipsoidal shells vs. volumes --  means that otherwise derived convergence criteria cannot be generally adopted. 

Although in this Section we have considered only Illustris-Dark results in order to focus on the convergence of the shape calculation with the number of particles in a halo, it might also be interesting to examine how halo shapes in Illustris vary with resolution. Such a result is necessarily affected by changes in the subgrid physics due to resolution and is further discussed in Appendix \ref{sec:IllustrisReso}. Briefly, we find larger deviations between the lower-resolution and high-resolution runs in comparison to the Illustris-Dark case. In fact, deviations persist at all halocentric radii: this is due to the fact that different resolutions imply slightly different resulting galaxy stellar masses, hence different star-formation efficiencies and hence different baryonic effects \citep[see Appendix 1 of ][]{Pillepich2018a}. However, a broad consistency between simulated and observed galaxies has been verified (and shall be intended) for the highest-resolution run Illustris: the effects of baryons on DM halo shapes from Illustris, and not from Illustris-2 or Illustris-3, are the ones that shall be considered the predictions from the Illustris galaxy-physics model.

\begin{figure}
	\centering
	\includegraphics[width=0.47\textwidth]{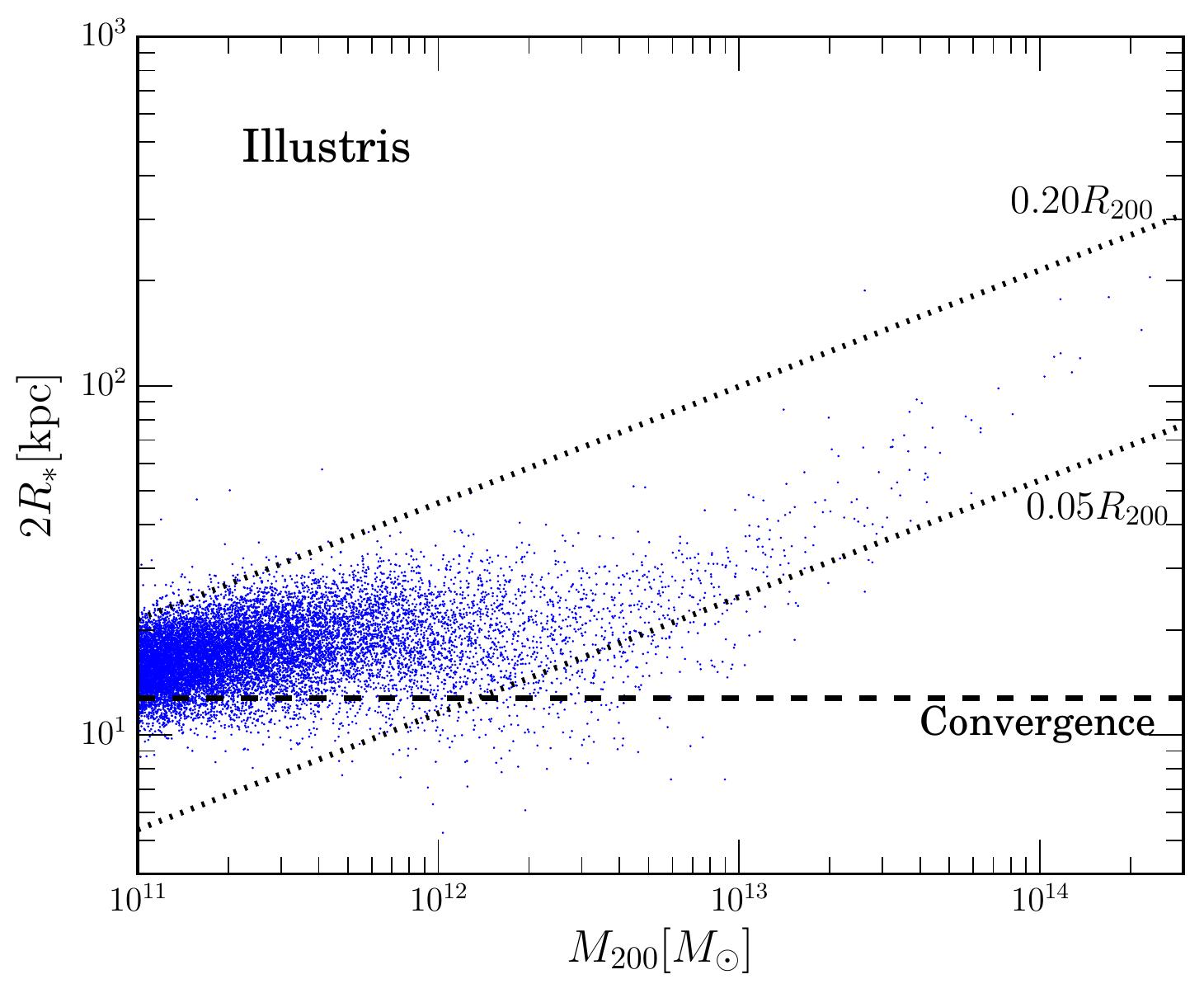}
	\caption{
		Galaxy size -- halo mass relationship in Illustris. Galaxy size is denoted by twice the stellar half-mass radius $(2R_*)$. The horizontal dashed line shows the minimum resolved radius $9\epsilon = 13$ kpc in this high resolution run. The upper and lower dotted lines show the radii corresponding to 20 per cent and 5 per cent of the halo virial radii. The majority of galaxy sizes lie within 20 per cent of their halo virial radii.
	}
	\label{fig:radiimass}
\end{figure}

Finally, before showing our results, we consider how different radial scales compare among each other for the considered Illustris haloes. Figure~\ref{fig:radiimass} shows in blue the galaxy size (defined here as twice the stellar half-mass radii or 2$R_*$) as a function of halo mass, for haloes of mass $>10^{11} \msun$ in Illustris. For comparison, the lines corresponding to $0.05R_{200}$ and $0.20R_{200}$ are also shown. We find that galaxies are typically contained with within 20 per cent of its halo virial radius. The horizontal dashed line in Figure~\ref{fig:radiimass} shows the minimum convergence radius for halo shapes in Illustris (13 kpc). Given the results of this section, in general, we will show only converged shape profiles i.e. for $r \gtrsim 9 \epsilon$. In fact, for the great majority of the haloes studied in this paper, this limit falls well inside our reference choice of `inner halo': $0.15 R_{200}$ (see Section~\ref{sec:inner}).

\begin{figure*}
	\centering
	\includegraphics[width=\textwidth]{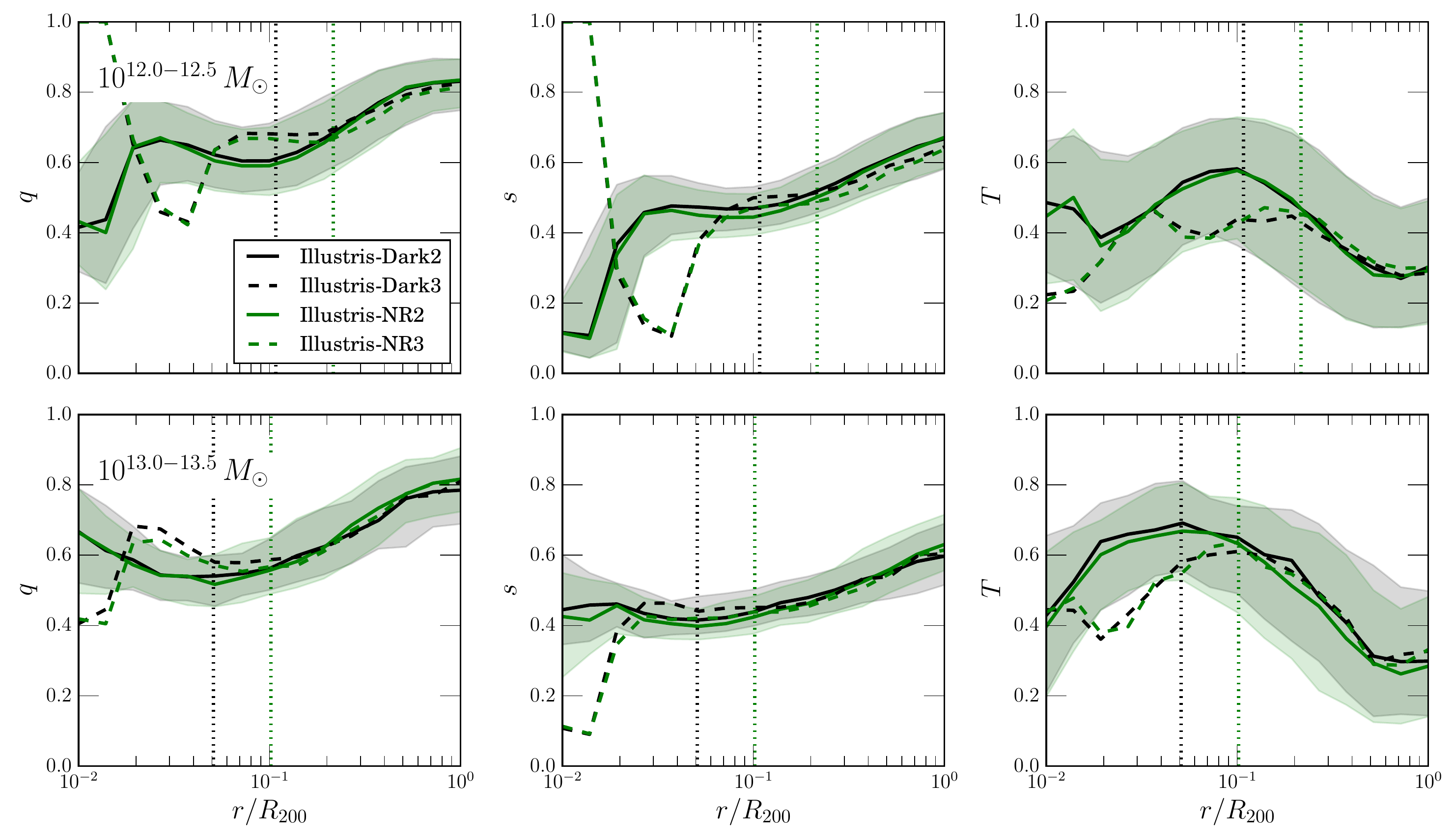}
	\caption{Comparison of median DM halo shape parameters $q\equiv b/a$ (left), $s\equiv c/a$ (middle) and $T\equiv (1-q^2)/(1-s^2)$ (right) as a function of radius in the DMO Illustris-Dark-2/3 and the non-radiative runs Illustris-NR-2/3.
		Results from haloes of mass $10^{12}$ are shown in the top row, while halo masses of $10^{13} M_\odot$ is shown in the bottom row.
		Black and green lines represent results from Illustris-Dark and Illustris-NR respectively.
		Vertical dotted lines show our convergence criteria for halo shapes.
		The moving-mesh hydrodynamics in Illustris-NR does not induce any changes in the halo shape when baryonic physics such as radiative cooling, star formation and feedback is turned off.
	}
	\label{fig:NR}
\end{figure*}

\section{Effects of Baryons on DM Halo Shapes}
\subsection{DMO and non-radiative halo shapes}

We show in Figure~\ref{fig:NR} the median shape parameters as a function of radius for Illustris-Dark (black) and Illustris-NR (green) for our two lower resolutions.
We find that the Illustris-Dark and Illustris-NR results are identical, thus non-radiative hydrodynamics alone does not induce any change in halo shapes. In the absence of any radiative processes, the gas neither cools and forms stars, nor is heated up by feedback processes. As a result, the present gas evolves similarly to the DM. 

In both Illustris-Dark and Illustris-NR, we find that haloes are least spherical near the halo centre, with axis ratios $\left<q \right> \approx 0.6$ and $\left<s \right> \approx 0.4$ at $r=0.15R_{200}$. 
Haloes become much more spherical near the virial radius, with axis ratios $\left<q \right> \approx 0.8$ and $\left<s \right> \approx 0.6$.
On the other hand, the triaxiality decreases towards the virial radius.
Hence, haloes are prolate near the halo centre and become more oblate with increasing radius.
These results are consistent with well-known results from other $N$-body studies of the halo shape.

\subsection{Radial Dependence}

\begin{figure}
	\centering
	\includegraphics[width=0.47\textwidth]{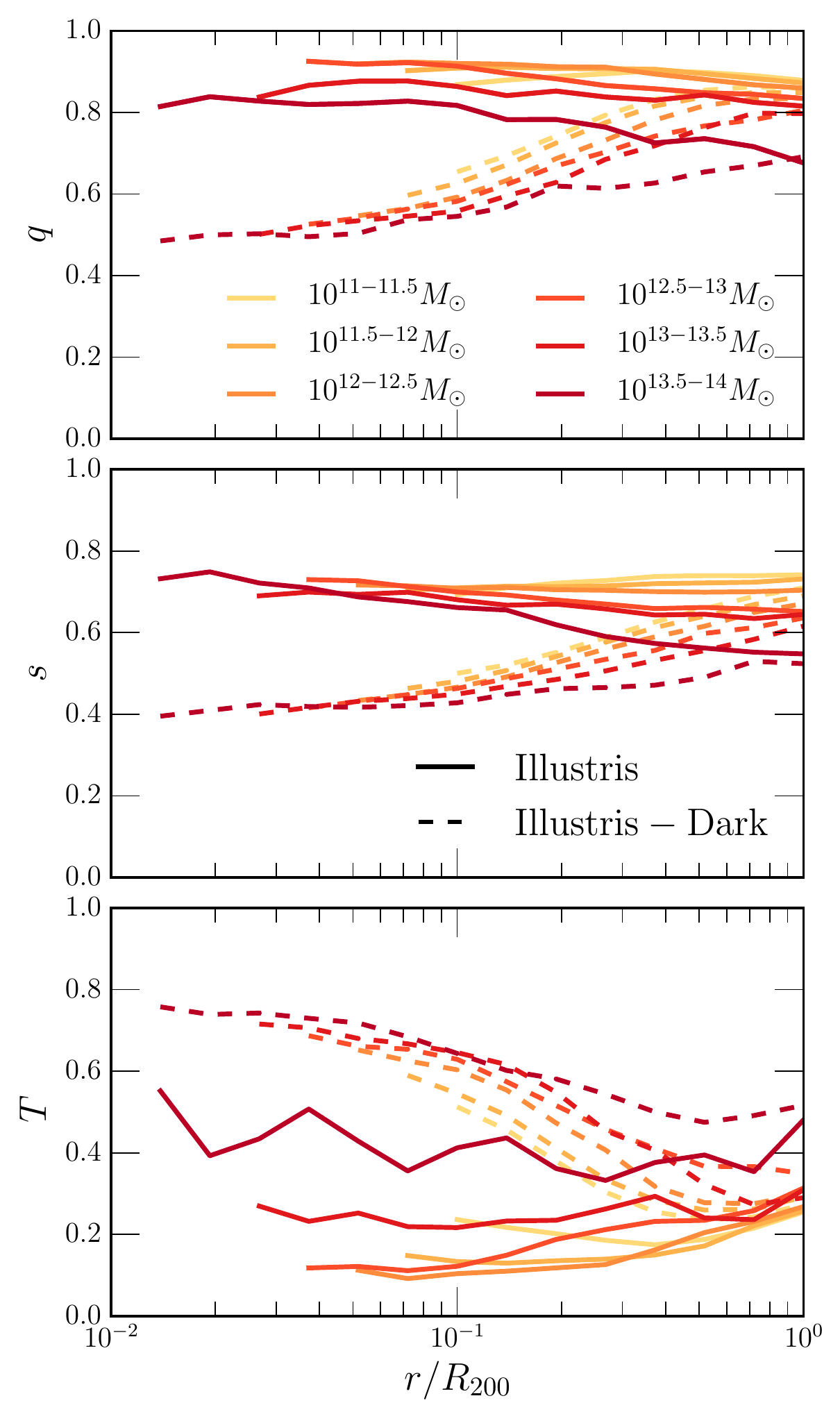}
	\caption{DM halo shape profiles for parameters $q\equiv b/a$ (top), $s\equiv c/a$ (middle) and $T\equiv (1-q^2)/(1-s^2)$ (bottom; $a>b>c$) as a function of halocentric distance. Only radii above the resolution limit of $\sim 9\epsilon$ have been shown.
	Solid and dashed lines represent results from the hydrodynamic simulation Illustris and the DMO simulation Illustris-Dark respectively.
	Colours denote different halo mass bins between  $10^{11} M_\odot$ and $10^{14} M_\odot$. For a given radius, baryons significantly sphericalise DM (increased $q$ and $s$) and make haloes more oblate (decreased $T$).
	This effect is strongest in the inner regions of haloes and becomes negligible towards the virial radius. 	
	}
	\label{fig:shape_all}
\end{figure}

The overall effects of baryons in Illustris can be seen in Figure~\ref{fig:shape_all}, where we plot the median shape parameters $q$ (top), $s$ (middle) and $T$ (bottom), together with the 25th and 75th percentile values of the halo population, as a function of halocentric distance.
The haloes are selected in six mass bins between $10^{11} M_\odot$ (light) and $10^{14} M_\odot$ (dark), while solid and dashed lines represent results from Illustris and Illustris-Dark, respectively.

The increase in axis ratios $q\equiv b/a$ (top) and $s\equiv c/a$ (middle) going from Illustris-Dark to Illustris shows that for a given radius, baryonic physics causes the DM halo to become significantly rounder. This effect is present throughout the halo, being strongest near the halo centre and decreasing towards the virial radius $R_{\rm 200}$. 
Coupled with the increase in $q$ and $s$, the triaxiality $T$ is also observed to decrease across all radii, indicating that haloes are more oblate at a given radius in Illustris compared to Illustris-Dark.

In both runs, we find that the shapes of DM haloes are generally not constant, but in fact vary with radius, albeit much more weakly in Illustris than Illustris-Dark. In Illustris-Dark, the DMO trend is for haloes to become more spherical and oblate towards the virial radius, which is consistent with previous $N$-body studies \cite[e.g.][]{Allgood06v367,Hayashi07v377}. On the other hand, above the convergence radius, in Illustris we find the variation with radius to depend on the halo mass: below $10^{12.5} \msun$, the axis ratios are almost independent of radius. Above $10^{12.5} \msun$, the axis ratios are found to decrease weakly with radius, with increasing steepness for more massive haloes. The triaxiality increases with radius in general, so Illustris haloes tend to become more prolate towards the virial radius. 

Our Illustris results are consistent with the smaller volume simulations of \cite{Abadi10v407} and \cite{Zhu17v466}, who also found the halo axis ratios to be roughly independent of radius for $10^{12} \msun$ haloes. This similarity occurs despite the absence of stellar and AGN feedback in \cite{Abadi10v407}, which accentuates the effect of baryons. On the other hand, using the MassiveBlack-II simulation, which has a similar box size and mass resolution to Illustris, \cite{Tenneti15v453} found  DM shapes to be flatter in the inner regions of haloes, with steeper profiles at lower masses. The contrasting results of MassiveBlack-II (MBII) and Illustris are likely a result of their differing baryonic physics implementations, which can be also seen in the ratio between FP and DMO halo masses for the two different simulations: the FP to DMO halo mass ratio is monotonic in MBII but non-monotonic in Illustris \citep{Chua2017}.

We note that in general, our results are not quantitatively comparable with previous studies on the radial dependence of DM halo shapes due to the different methodologies that have been employed to infer halo shapes. For example, both \cite{Allgood06v367} and \cite{Tenneti15v453} relied on the iterative reduced inertia while \cite{Abadi10v407} and \cite{Kazantzidis10v720} inferred halo shape profiles by approximating the iso-potential surfaces with ellipsoids. 


\subsection{Defining the inner and outer haloes}
\label{sec:inner}

\begin{figure*}
	\centering
	\includegraphics[width=\textwidth]{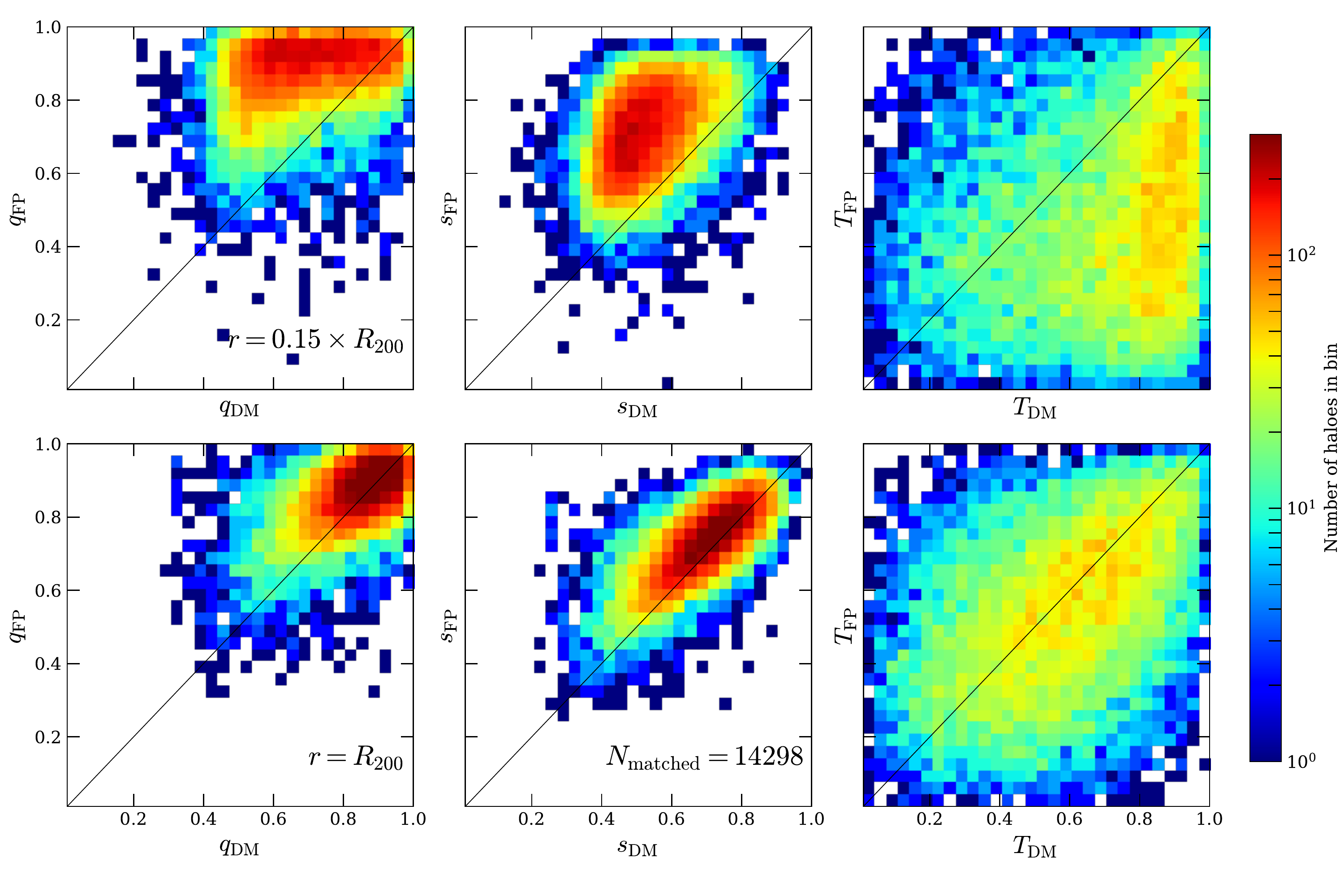}
	\caption{2D histogram of DM halo shape parameters in Illustris against the matched haloes in Illustris-Dark, for all halo with mass $M_{200} > 10^{11} M_\odot$ (14298 matched halo pairs in total). Bottom and top rows show the results for $r=0.15R_{200}$ and $r=R_{200}$ respectively. In Illustris, the inner halo ($0.15R_{200}$) is significantly more spherical and oblate than the Illustris-Dark counterparts. At the virial radius, the effect of baryons is negligible, and the shape parameters are well correlated between Illustris and Illustris-Dark, yet with some non-negligible scatter.
	}
	\label{fig:corr_dmfp}
\end{figure*}

To better understand how the shape of individual haloes are changed, we investigate halo shapes at fixed fractions of the virial radius.
Since the effect of baryons is not uniform with radius, we measure the shapes of the inner and outer halo, separately:
\begin{enumerate}
	\item outer halo shape: the local shape at the virial radius $R_{200}$
	
	\item inner halo shape: the local shape at $R_{15} \equiv 0.15 R_{200}$. 
\end{enumerate}

The choice of $R_{15}$ is motivated by observational measurements of the Galaxy, which is restricted to the regions relatively near the halo centre or close to the Sun. For example, \cite{Law10v714} measured the MW shape at a range of 16-60 kpc from the galactic centre. Since the MW has a virial radius of $R_{200} \approx 200{\rm kpc}$, this corresponds to $R_{15} \approx 30{\rm kpc}$, lying within the \cite{Law10v714} study.
While it is advantageous to measure the halo shape close to the halo centre where baryonic effects are most pronounced, our choice of $R_{15}$ is further guided by the convergence studies of Section \ref{sec:resolution}. We find that $R_{15} >$ 12 kpc for haloes of mass $M_{200} > 10^{11} M_\odot$, thus the inferred halo shapes are well converged at this radius.

\subsection{Quantifying the effects of baryons in the inner and outer haloes}

Figure~\ref{fig:corr_dmfp} plots the 2D histograms of halo shapes by showing the correlation between the shape parameters of Illustris and Illustris-Dark for all matched haloes with $M_{200} > 10^{11} M_\odot$. 14298 such pairs were identified between the two runs. Diagonal black lines represent the 1:1 case where the DM shapes in Illustris are unchanged from that in Illustris-Dark. At an inner radius of $r=0.15R_{200}$ (top row), both $q_{\rm FP}$ and $s_{\rm FP}$ are highly boosted from their Illustris-Dark values, signifying their increased sphericities.

More importantly, we find that the shape parameters remain correlated to their matched DMO counterparts: haloes which are more spherical remain more spherical in Illustris-Dark as well. This suggests that, while baryonic physics impact shapes significantly, their effects continue to depend, most probably, on other halo properties such as formation time and concentration -- see next Sections.

At the virial radius $R_{200}$, the bottom row of Figure~\ref{fig:corr_dmfp} indicates a much weaker effect of baryons.
At this radius, both the normalization and gradient of the shape parameters in Illustris remain close to their matched Illustris-Dark counterparts. These statements hold for the bulk of the halo population, as in fact there are cases where halo shapes are completely different between the FP and DMO runs. In other words, the scatter in the plots of Figure~\ref{fig:corr_dmfp} is not negligible.

\begin{figure*}
	\centering
	\includegraphics[width=\textwidth]{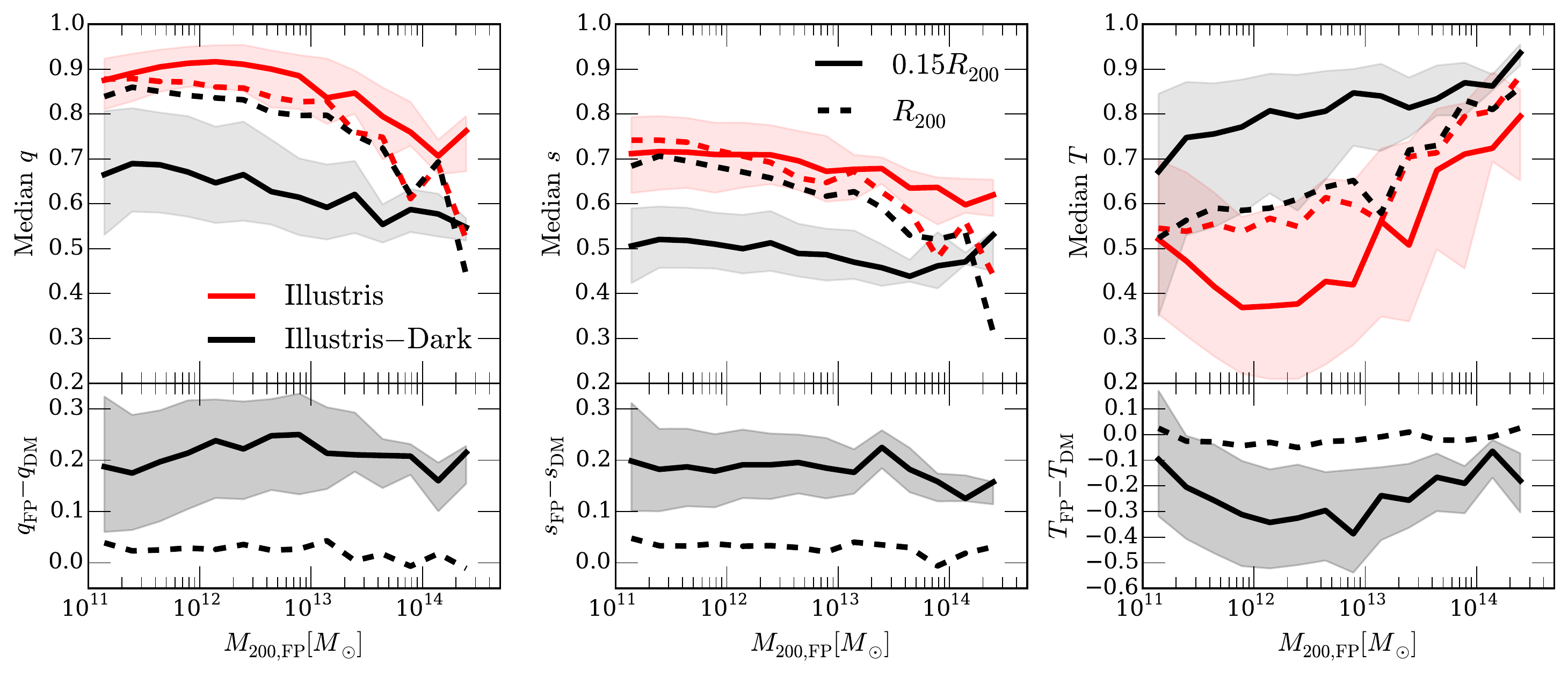}
	\caption{Top row: Median DM halo shape parameters for matched haloes as a function of halo mass, calculated at $r=0.15R_{200}$ (solid lines). The shaded regions denote the 25th and 75th percentile of the galaxy/halo distributions. For comparison, results calculated at $R_{200}$ are also shown as dashed lines. Red and black lines correspond to results from Illustris and Illustris-Dark respectively. 
        Bottom row: Difference between Illustris and Illustris-Dark shape parameters as a function of halo mass. Median values of $q$ and $s$ are boosted by about 0.2 almost across halo masses in Illustris compared to Illustris-Dark.
	}
	\label{fig:shape_m200}
\end{figure*}

We plot in Figure~\ref{fig:shape_m200} the median shape parameters at $0.15R_{200}$ as a function of halo  mass for Illustris (red) and Illustris-Dark (black) as well as the difference between the two runs. In $N$-body studies, the halo mass is an important halo property, correlating well with parameters such as the formation time, concentration, subhalo abundance and spin \citep[e.g.][]{Jeeson11v415,Skibba11v416}. Halo shapes have also been found to correlate well with mass, and numerical simulations point to a negative correlation of the median sphericity $\left<s\right>$ with halo mass. A parametrization of the sphericity--mass relation is given in \cite{Allgood06v367}, which found $\left<s\right>$ to be well-described by a simple power law $\left<s_{0.3}\right>=a(M_{\rm vir}/M_{*}(z,\sigma_8))^{b} $ where $s$ is measured inside  $0.3R_{\rm 200}$, $M_{\rm vir}$ is the virial mass of the halo and $M_*(z,\sigma_8)$ is the characteristic non-linear mass for the cosmology and redshift, with fitting parameters $a$ and $b$ \footnote{They found the following values for the fitting parameters:	$a = 0.54 \pm 0.02$ and $b = -0.050 \pm 0.003$. An alternative parametrization given in \cite{Maccio08v391} is $\left<s_{0.3}\right> = c + d \log_{10}(M_{\rm vir}/M_{*})$ for fitting parameters $c$ and $d$.}. 

In Illustris-Dark, both $\left<q\right>$ and $\left<s\right>$ anti-correlate with and decrease monotonically with mass, albeit not very strongly, in agreement with previous $N$-body simulations \citep{Allgood06v367,Maccio08v391,Butsky16v462}.
In Illustris, the anti-correlation with mass is overall retained, and at the same time, the primary effect of baryons at $0.15R_{200}$ is to increase the median $q$ and $s$ by $\approx +0.2$ and $T$ by $\approx -0.3$: this means increased sphericity and oblateness of the inner halo. Again, there is negligible difference between the two runs at virial radius. In Table \ref{table:shapemass}, we provide fitting parameters for $\left<q \right>$ and $\left<s \right>$ in the form of $\left<q,s\right> = a(M_{\rm vir}/10^{12}\msun)^{b} $ at three different radii: $0.15R_{200}$, $0.3R_{200}$ and $R_{200}$.

\begin{table}
	\centering
	\begin{tabular*}{0.4\textwidth}{@{\extracolsep{\fill}}l c  c | c  c |  c  c}
		\hline
		& \multicolumn{2}{c|}{$0.15 R_{200}$}	& \multicolumn{2}{c|}{$0.3R_{200} $} & \multicolumn{2}{c}{$R_{200}$} \\ 
		& $a$	&				$b$	& $a$	&  $b$ 	& $a$	& $b$\\
		\hline
		$q_{\rm FP}$	& 0.87	& -0.027 & 0.86 & -0.035 &  0.85 & -0.059\\ 
		$s_{\rm FP}$	& 0.70	& -0.024 & 0.70 & -0.039 & 0.71 & -0.072 \\  
		\hline
		$q_{\rm DMO}$	& 0.68	& -0.036 & 0.76	& -0.041 & 0.82 & -0.058\\
		$s_{\rm DMO}$	& 0.52	& -0.022 & 0.58	& -0.042 & 	0.67 &-0.070\\
		\hline
	\end{tabular*}
	\caption{
		Fitting parameters to the equation $\left<p\right>=a~(M_{\rm 200}/10^{12}\msun)^{b}$ in Illustris (FP) and Illustris-Dark (DMO) for three different radii, with $p \equiv q$ or $s$. Results at $0.3R_{200}$ are provided for comparison with Allgood et al. (2006).
	}
	\label{table:shapemass}
\end{table}

In addition to the overall negative correlation with halo mass, our Illustris results also exhibit a secondary effect which breaks the monotonicity of the relations observed in Illustris-Dark. We find in Illustris that the parameters $q$ and $T$ peak and dip respectively between a halo mass of $10^{12} - 10^{13} M_\odot$. A similar trend is visible also in the bottom row where we plot the difference in the values of the parameters between matched haloes in Illustris and Illustris-Dark.

\begin{figure*}
	\centering
	\includegraphics[width=\textwidth]{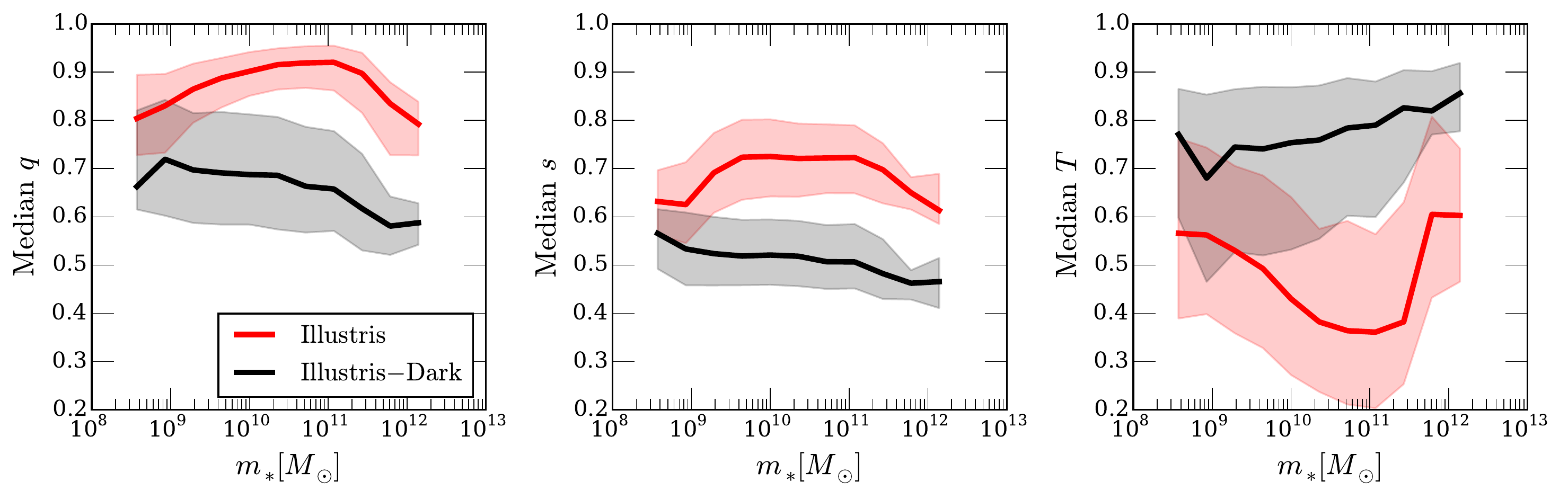}
	\caption{Median DM halo shape parameters for matched haloes as a function of stellar mass, measured at $r=0.15R_{200}$ (solid lines). Results from Illustris and Illustris-Dark are shown as red and black lines respectively. The shaded region denotes the 25th and 75th central quartiles. Illustris haloes with stellar masses of $\approx 10^{10.5-11} M_\odot$  have the most spherical and oblate inner haloes. The shapes of corresponding matched haloes in Illustris-Dark do not exhibit such a trend.}
	\label{fig:shape_mstar}
\end{figure*}

The non-monotonicity of the inner halo shape as a function of mass in Illustris is more evident using stellar mass instead of halo mass. Figure~\ref{fig:shape_mstar} plots the shape parameters as a function of stellar mass, which we measure within twice the stellar half-mass radius. Median results from Illustris and the corresponding matched haloes in Illustris-Dark are shown as red and black solid lines respectively, with shaded region showing the 25th to 75th central quartiles of the galaxy population. Here, Figure~\ref{fig:shape_mstar} shows clearly the non-monotonic behaviour that was alluded to in Figure~\ref{fig:shape_m200}. In particular, the parameters $q$ and $T$ have a peak and trough respectively at $m_* \approx 10^{11} \msun$, showing that these haloes of these stellar masses are most spherical and oblate in Illustris. Again, the matched haloes from Illustris-Dark do not exhibit such a behaviour, showing that the non-monotonic modification of the shape is a direct result of baryonic physics, and not a secondary reflection of other halo properties.  The difference between Figures \ref{fig:shape_mstar} and \ref{fig:shape_m200} can be explained by scatter in the stellar mass -- halo mass relation, which suppresses the peak when halo mass is used.

\subsection{Effect of baryons on velocity anisotropy}

\begin{figure*}
	\centering
	\includegraphics[width=\textwidth]{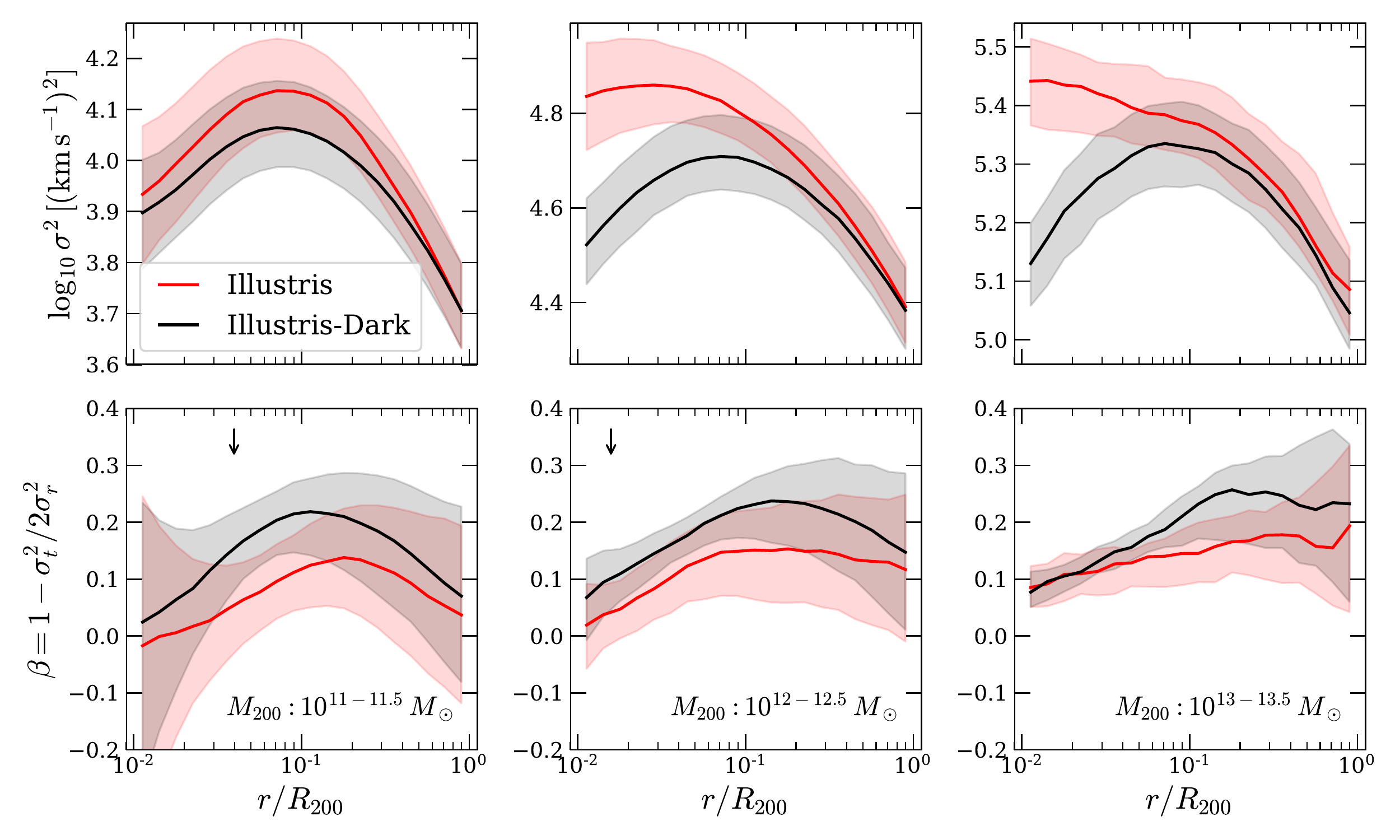}
	\caption{Effects of baryons on the velocity structure of DM in haloes of mass $10^{11} \msun$ (left), $10^{12} \msun$ (middle) and $10^{13} \msun$ (right). Top row and bottom row show the median velocity dispersion and  the median velocity anisotropy ($\beta$) as a function of radius respectively. Solid lines correspond to the median while shaded area denotes the 25th to 75th percentiles. Results for Illustris and Illustris-Dark are shown in red and black respectively. Arrows denote the P03 convergence radii in Illustris-Dark. Illustris haloes exhibit larger velocity dispersions (especially in the inner halo) and are more isotropic (smaller $\beta$) compared to Illustris-Dark.
	}
	\label{fig:beta}
\end{figure*}

The velocity dispersion structure of DM haloes, defined as $\sigma^2 = \left< \left( v - \left< v \right>\right)^2 \right>$, has been studied in previous $N$-body simulations \citep[e.g.][]{Navarro2010} and hydrodynamic simulations \citep[e.g.][]{Pedrosa2010,Tissera10v406}. The $N$-body results showed that DMO haloes show a temperature inversion near the centre where the velocity dispersion decreased at small radii. We calculate the halo velocity dispersion profiles $\sigma^2(r)$ in spherically symmetric shells of radius $r$. The top row of Figure~\ref{fig:beta} compares the total velocity dispersion profiles of Illustris (red) and Illustris-Dark haloes (black). On average, we find that baryons increase the velocity dispersions, especially in the central regions. The increased central velocity dispersion results in dispersion profiles that decrease monotonically with radius for $10^{12}$ and $10^{13} \msun$ haloes, as reported in previous hydrodynamic work on galaxy-sized haloes \citep{Pedrosa2010,Tissera10v406}. The velocity dispersion of $10^{11} \msun$ haloes remain non-monotonic in spite of the increased central velocity dispersion.

\begin{figure}
	\centering
	\includegraphics[width=0.47\textwidth]{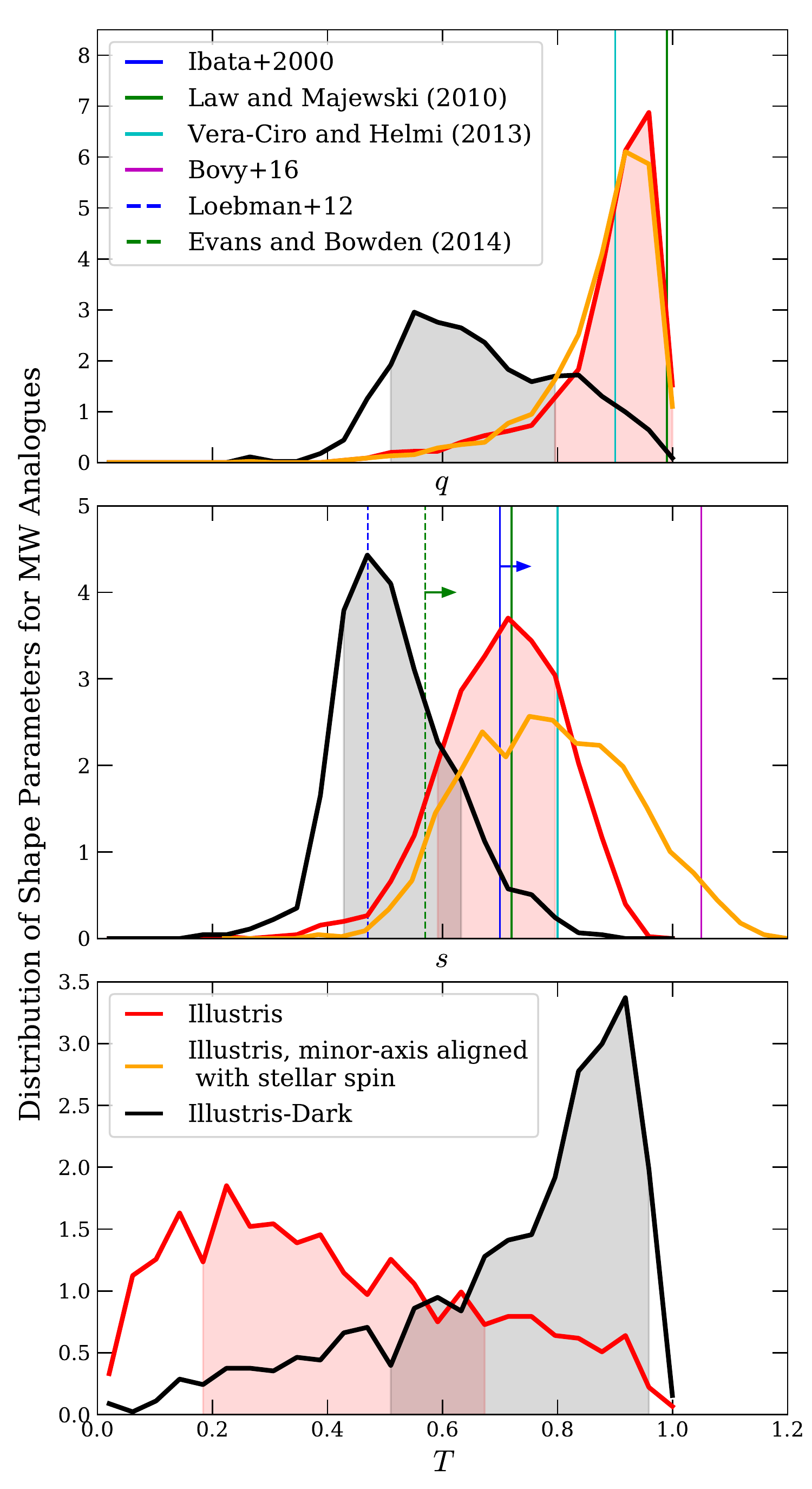}
	\caption{Comparison of simulated MW-analogues in Illustris and Illustris-Dark with observations of the MW halo shape. We plot the distribution of the inner halo ($r=0.15R_{200}$) shape parameters for haloes of mass $8\times 10^{11} - 2 \times 10^{12} M_\odot$ for both Illustris (red) and Illustris-Dark (black). The orange distributions show the fixed-axis parameters (for $q$ and $s$ only), where the halo minor axis is constrained to lie along the direction of stellar spin, as is the case for certain observational results. The vertical lines show various measurements derived from observations of MW stellar streams (solid lines) and stellar kinematics (dashed lines). Arrows on observations denote lower bounds.
	}
	\label{fig:MWshape}
\end{figure}

The bottom row of Figure~\ref{fig:beta} compares the median velocity anisotropies ($\beta$) of haloes in Illustris and Illustris-Dark, which summarizes the relative abundance of radial and circular orbits of th DM particles. In general, we find haloes to be most isotropic ($\beta \approx 0$) near the central regions, become more radially-biased ($\beta > 0$) at larger radii before becoming more isotropic again near the virial radius. Baryons alter the orbital structure by decreasing the dominance of radial motions. Unlike for halo shapes, where the Illustris and Illustris-Dark distributions are well-separated (see e.g. Figure~\ref{fig:shape_m200} or Figure~\ref{fig:MWshape}), there is substantial overlap between the velocity anisotropies of the two runs. As \cite{Tissera10v406} found from hydrodynamic re-simulations of the Aquarius haloes, baryonic effects can vary dramatically between individual haloes. For example, they found that only three of their haloes become less radially dominated, while the other three remain similar to their DMO counterparts. The lack of baryonic effects on the velocity anisotropy of some haloes would explain the small separations of the two runs and is consistent with the large scatter in halo shapes between FP and DMO analogue haloes of Figure~\ref{fig:corr_dmfp}.


\section{A closer look into Milky Way-sized haloes}

In the previous Section, we have found that halo shapes depend on halo and galaxy properties such as the halo and stellar mass. While the halo mass is often identified as an important halo property in $N$-body simulations, other properties such as the halo formation time, concentration and spin can be also fundamental in determining halo shapes \cite[e.g.][]{Jeeson11v415}. In this section, we examine the relation between halo shape and other fundamental halo properties to understand what drives halo shapes both in $N$-body as well as  hydrodynamic simulations. However, we focus here on MW-mass haloes.

\subsection{Comparison with Milky-Way observations}
Before looking into other halo properties, we first turn to MW analogues in our simulations to understand how the shapes of simulated MW-like haloes are distributed, and also to compare our results with observations. 

Currently, the best measurements of halo shapes come from the MW, since the motion of individual stars can be resolved and measured. One method of inferring our Galaxy's shape uses stellar kinematics (measured by e.g. SDSS) for equilibrium modelling with the Jeans equations \citep[e.g.][]{Loebman2012,Bowden16v460}. Another class of methods uses stellar streams formed from the tidal stripping of satellite galaxies or globular clusters. These include the measurements of \cite{Ibata01v551}, \cite{Law10v714}\footnote{As many studies have pointed out, \cite{Law10v714} measured the major axes of the halo to be in the plane of the disc. Such intermediate-axis orientations have been found to be unstable in numerical modelling of disk galaxies \citep{Debattista13v434}. As with other numerical simulations, we find in Illustris a preference for the minor axes of the halo and the disk to be aligned.} and \cite{Vera-Ciro13v773} which were made using the tidal tails of the Sgr dwarf galaxy, and of \cite{Bovy16v833} which were made using the Pal 5 and GD-1 tidal streams. Because these measurements rely on halo stars and tidal streams, they are limited to the inner halo where these stars reside and can be observed. Here, we compare the results of these observations to the MW analogues we find in our simulations.

In order to compare our results with the afore-mentioned observations, we first note that \cite{Bovy16v833} reported the MW halo shape assuming the halo minor axis to be aligned with with that of the stars, or in other words, perpendicular to the MW disk, if a disk is in place. This differs from the iterative method described in Section \ref{sec:iterative} which places  no such restriction on the DM axes. Consequently, we denote the parameter $s_{\rm fixed}=c'/a'$ as the flattening perpendicular to the stellar disk, and $q_{\rm fixed}=b'/a'$ as the parameter describing axi-symmetric deviations in the disk plane. Misalignments between the stellar and DM shapes, as noted in various cosmological hydrodynamic simulations, result in to $s_{\rm fixed} \neq s$ and $q_{\rm fixed} \neq q$. In Illustris, \cite{Tenneti16v462} has found a substantial misalignment between the stars and the DM halo: the mean 3D misalignment angle between the major axis of the stars and the DM halo was found to be $\approx 46^o$ for disk and $\approx 37^o$ for elliptical galaxies. 

To derive $q_{\rm fixed}'$ and $s_{\rm fixed}'$ for our simulations, we impose the requirement that the minor axis $z'$  be parallel to the stellar disk spin. The $x'$ and $y'$ axes thus lie in the plane of the disk. First, we associate the stellar minor axis with the stellar disk spin, defined as ${\mathbf j}^* = \left(\sum_{i} m_i {\mathbf r_i} \times {\mathbf v_i}\right)/\sum_i m_i$, where the summations involve stellar particles contained within twice the stellar half-mass radius $(r<2r_{1/2})$. 
Then, starting with the converged shape tensor $S_{ij}$ from the iterative procedure described in Section \ref{sec:iterative}, we rotate the shape tensor into a frame where the $z'$-axis is aligned with the stellar disk spin ${\mathbf j}^*$. In the rotated primed frame, $\lambda_c'$ is taken to be the component of the rotated shape tensor $S'_{ij}$ lying along the $z'$-axis. In the plane of the disk, i.e. the $x'$ and $y'$ directions, the shape tensor is a $2\times 2$ matrix $S'_{ij}$ which is diagonalised to obtain the eigenvalues $\lambda_a'$ and $\lambda_b'$. As before, the axis ratios are determined using the square roots of the eigenvalues: $q_{\rm fixed}=\sqrt{\lambda_b'/\lambda_a'}$ and $s_{\rm fixed}=\sqrt{\lambda_c'/\lambda_a'}$. We denote these derived parameters as the \emph{fixed-axis} parameters, which are mainly used for comparisons with the  \cite{Bovy16v833} results. We do not distinguish between galaxy morphologies, since we do not find a significant difference even when morphological differences are considered.

Figure~\ref{fig:MWshape} shows the distribution of shape parameters of MW analogues (halo mass $8\times 10^{11} - 2 \times 10^{12} M_\odot$) in the inner halo ($r=0.15R_{200}$) of Illustris (red) and Illustris-Dark (black), together with the afore-mentioned observational measurements that have been made of the MW halo shape (vertical lines). Note that, because of the radial-independence of shape parameters at the MW-mass scale, it does not matter to what galactocentric distances our results are quoted, at least in Illustris. The orange distributions correspond to the fixed-axis shape parameters where the DM minor axis is restricted along the stellar disk spin.
For MW analogues in Illustris, we find that 
$q_{\rm FP}  = 0.88 \pm 0.10$ and $s_{\rm FP}  = 0.70 \pm 0.11$ compared to 
$q_{\rm DMO}  = 0.67 \pm 0.14$ and $s_{\rm DMO}  = 0.52 \pm 0.10$ for Illustris-Dark. 
These 1-$\sigma$ intervals are represented by shaded regions in Figure~\ref{fig:MWshape}. The large shifts between the Illustris and Illustris-Dark distributions are again results of the sphericalisation by baryons.
The high value of $q_{\rm FP}$ (close to unity) indicates that the Illustris haloes are close to, but not completely axisymmetric. Observations of the azimuthal abundance of MW disk stars \citep{Bovy2014} near the Sun as well as their kinematics \citep{Bovy2015}, constrain the halo axis ratio $q$ to be close to unity in the inner halo, which is highly disfavored in the DMO Illustris-Dark. The Illustris (red) and the fixed minor-axis (orange) distributions are similar for the axis ratio $q$, indicating that halo misalignment does not appreciably affect its determination. For the axis ratio $s$, however, halo misalignment between stars and DM causes a noticeable shift towards larger values, and results in haloes appearing to be more spherical than if the stellar and DM shapes were allowed to be misaligned. We obtain on average $s_{\rm fixed} = 0.79 \pm 0.15$ in Illustris, when the halo minor axis is constrained along the direction of the stellar spin.

In Figure~\ref{fig:MWshape}, the solid vertical lines show the measurements made using stellar streams while dashed vertical lines are results from stellar kinematics, most of which have focused on the minor-to-major axis ratio $s$. Interestingly, these observational results seem to be discrepant with one another, with a large dispersion and with results for $s$ ranging from 0.5 to 1. 
With the exception of the \cite{Loebman2012} result, the Illustris haloes exhibit much stronger agreement with these observations than Illustris-Dark. 
The measurements using Sgr. dwarf \citep{Ibata01v551,Law10v714,Vera-Ciro13v773} are similar, and agree very well with the Illustris shape distributions, lying within 1$\sigma$ of the Illustris predictions.

The results of \cite{Bovy16v833} (magenta line) found a value of $s=1.05 \pm 0.14$ (2$\sigma$: 0.79--1.33) for the MW, thus favouring an extremely spherical halo.
Comparing their result with the Illustris fixed minor-axis results of Figure~\ref{fig:MWshape} (orange distributions), we find a combined uncertainty of $\sigma = 0.2$, Since this is smaller than the difference in the mean values ($\Delta s = 0.26$), the \cite{Bovy16v833} measurement is more spherical and thus disagrees with  the Illustris predictions at the 1$\sigma$ level.


\subsection{Correlation with Halo Properties}

\begin{figure*}
	\centering
	\includegraphics[width=\textwidth]{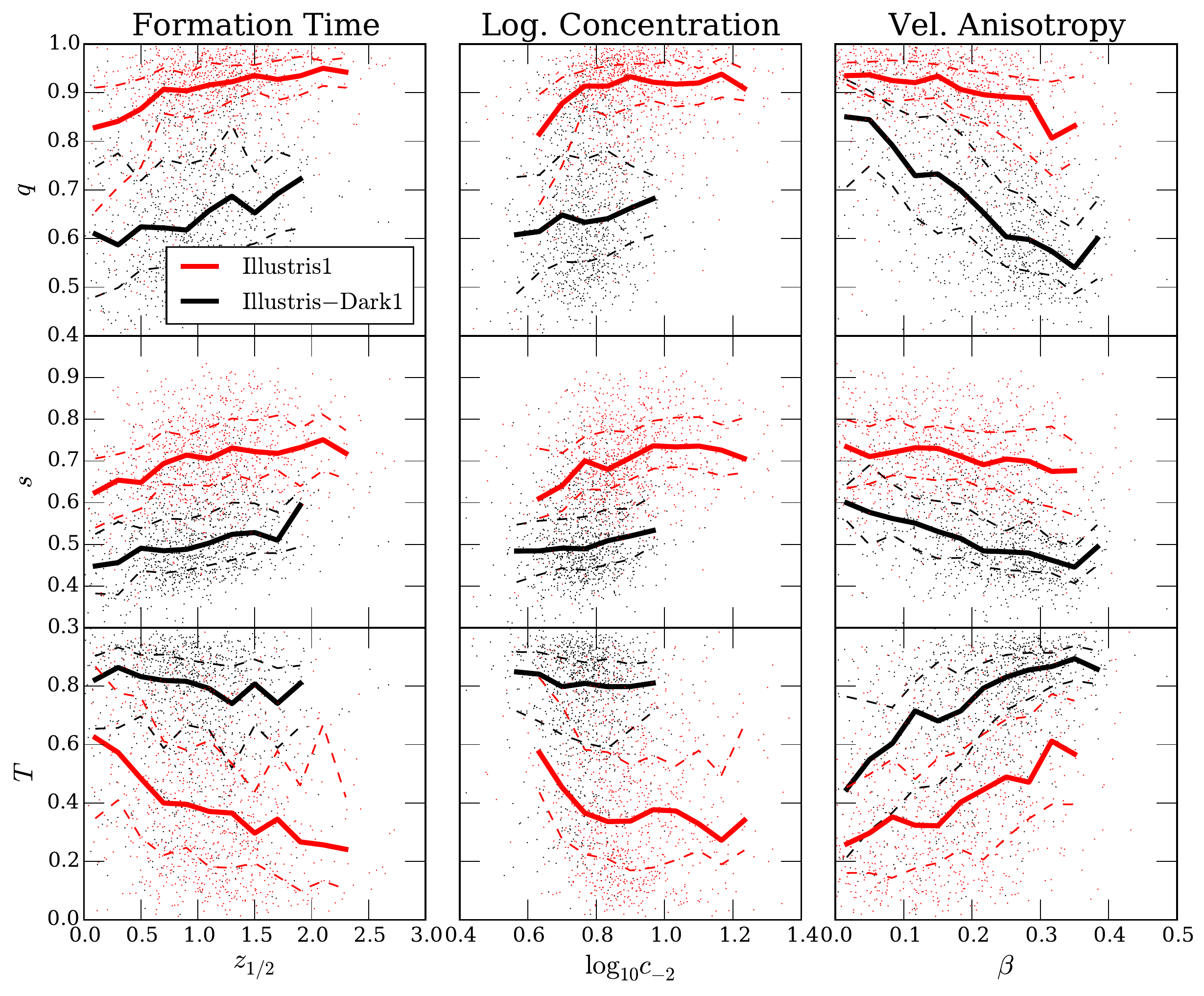}
	\caption{Correlation of DM halo shape parameters measured at $r=0.15R_{200}$ with the halo formation time (left), DM concentration parameter (middle) and the velocity anisotropy (right), for haloes of mass $10^{12-12.5} M_\odot$. Illustris and Illustris-Dark results are shown in red and black respectively. Solid lines indicate the median of the distribution, while dashed lines show the 25th and 75th central quartile. Spearman correlation statistics are shown in Table \ref{table:halocorr}. The velocity anisotropy parameter correlates most strongly with halo shape in Illustris-Dark, while all three properties correlate with halo shape in Illustris, to a smaller degree.
	}
	\label{fig:halocorr}
\end{figure*}

\begin{table}
	\centering
	\begin{tabular*}{0.47\textwidth}{@{\extracolsep{\fill}}l c   c    c  }
		\hline
        & \multicolumn{3}{c}{Spearman Correlation}\\
		& \multicolumn{1}{c}{$z_{1/2}$}	& \multicolumn{1}{c}{$ \log_{10} c_{-2}$} & \multicolumn{1}{c}{$\beta$} \\ 
		\hline
		$q_{\rm FP}$	& 0.29	 & 0.19 &   -0.28 \\ 
		$s_{\rm FP}$	& 0.22	 & 0.25 & 	-0.067 \\  
		$T_{\rm FP}$	& -0.23	 & -0.13 &  0.27 \\ 
		\hline
		$q_{\rm DMO}$	& 0.18	 & 0.093	& -0.49 \\
		$s_{\rm DMO}$	& 0.22	& 0.12	& 	-0.35 \\ 
		$T_{\rm DMO}$	& -0.11	& -0.056& 0.46 \\  
		\hline
	\end{tabular*}
	\caption{
		Spearman correlation values corresponding to Figure~\ref{fig:halocorr}. Correlation statistics are shown between the shape parameters and the halo properties: formation time $z_{1/2}$, halo concentration $\log_{10} c_{-2}$ and velocity anisotropy $\beta$. Illustris and Illustris-Dark results are denoted as FP and DMO respectively.
		}
	\label{table:halocorr}
\end{table}

We correlate the inner halo shape parameters with formation time, concentration, and velocity anisotropy parameter, in Figure~\ref{fig:halocorr}, with results from Illustris and Illustris-Dark shown in red and black, respectively. The solid line shows the  median values while dashed lines show the 25th and 75th central quartile. To better quantify the correlation, we calculate the Spearman correlation value $\rho$ in Table~\ref{table:halocorr}, which measures the monotonicity of relationship between the parameters. Correlations of -1 or +1 indicate exact monotonicity while $\rho=0$ indicates no correlation.

Table \ref{table:halocorr} shows that halo shape correlates most strongly with the velocity anisotropy parameter $\beta$ in both runs, with stronger correlations in Illustris-Dark compared to Illustris. In Illustris-Dark, $q_{\rm DMO}$ exhibits the strongest correlation with $\beta$, with a Spearman correlation value of $-0.49$. The strong correlation between halo shape and the velocity anisotropy arises because the shape of the collisionless DM halo has to be sustained by the velocity dispersion \citep{Allgood06v367}. In general, the axis ratios $q$ and $s$ anti-correlate with $\beta$, while $T$ correlates positively: haloes that are more dominated by circular orbits are both more spherical and oblate. Interestingly, the sphericity $s$ and $\beta$ do not correlate in Illustris. 

For the halo formation time, we find from Figure~\ref{fig:halocorr} similar trends between the two runs: haloes that form earlier are both more spherical and oblate. This is reflected in the Spearman correlation values which are positive with $q$ and $s$ and negative with $T$, consistent with previous $N$-body studies.

In contrast, the concentration parameter exhibits quite different behaviours in Illustris and Illustris-Dark.
In Illustris-Dark, the small Spearman correlations ($\left| \rho \right| \lesssim 0.1$) indicate very little correlation between halo shapes and  concentration.
Including baryon physics in the simulation boosts the correlations substantially, resulting in correlations similar to that of the formation time: 
Illustris haloes with larger concentrations are also more spherical and oblate. 

We also note that Figure~\ref{fig:halocorr} shows that, in addition to affecting halo shapes, baryons raise the halo concentration for $10^{12} \msun$ haloes in Illustris compared to Illustris-Dark. This was previously observed in Illustris in \cite{Chua2017} and is reflective of halo contraction that has been predicted theoretically and observed in some previous hydrodynamical simulations \citep[e.g.][]{Blumenthal86v301,Gnedin04v616,Duffy10v405}.

The lack of correlation between halo shape and concentration in Illustris-Dark is in contrast with the results of \cite{Jeeson11v415}, who found using a Principal Components Analysis (PCA)  study of $N$-body haloes that the concentration correlates well with the sphericity $s$. We believe this is due to (1) \cite{Jeeson11v415} defining the concentration using the NFW profile and (2) calculating the halo shape with a non-iterative method that is less accurate \citep{Zemp11v197}.

\subsection{Correlation with galaxy formation efficiency}
\label{sec:galform}

\begin{figure*}
	\centering
	\includegraphics[width=\textwidth]{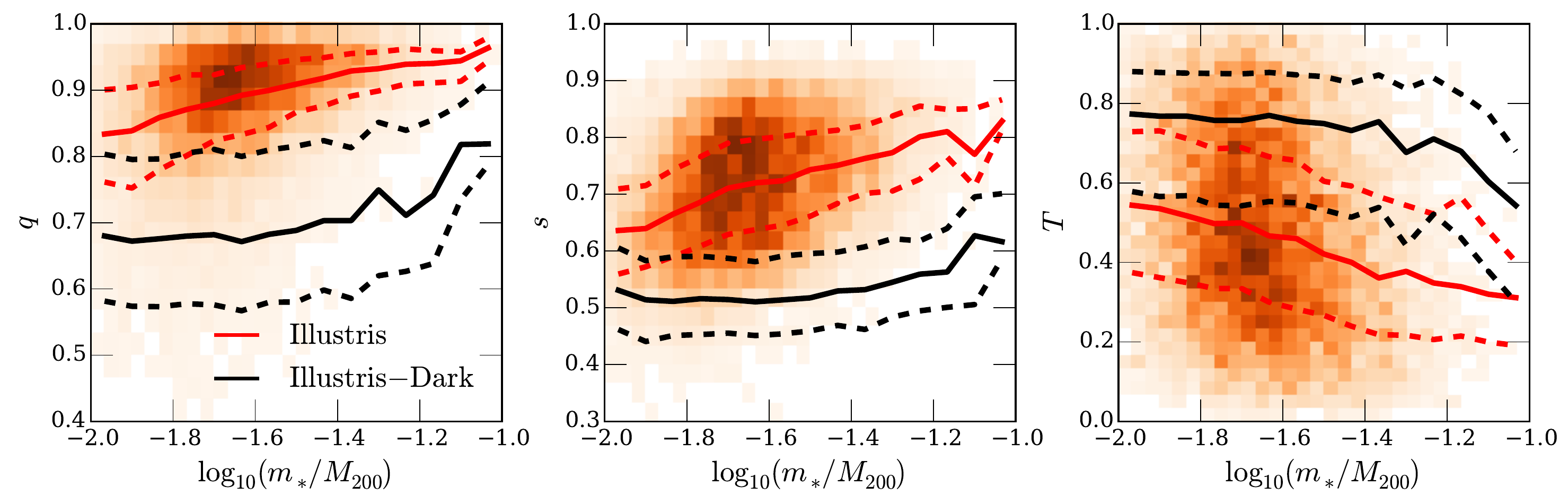}
	\caption{Dependence of DM halo shape parameters measured at $r=0.15R_{200}$ on the galaxy formation efficiency ($m_*/M_{200})$ for all haloes with mass $> 10^{11} M_\odot$.
		Results for Illustris-1 are shown in red.
		Black lines are shape parameters drawn from the matched halo in Illustris-Dark.
		Solid and dashed lines denote the median and the 25th to 75th percentile of the distribution.
		Contour plots in the background show the distributions of the shape parameters with galaxy formation efficiency for Illustris.
		The corresponding Spearman correlations are shown in Table \ref{table:galformcorr}. We find that the axis-ratio parameters strongly correlate with the stellar-to-halo mass ratio in Illustris.
	}
	\label{fig:galformcorr}
\end{figure*}

\begin{table}
	\centering
	\begin{tabular*}{0.45\textwidth}{ c c | c c}
		\hline
		& Spearman correlation	& & Spearman correlation		\\
		\hline
		$q_{\rm FP}$	& 0.34	& $q_{\rm DMO}$	& -0.055 \\ 
		$s_{\rm FP}$	& 0.33	& $s_{\rm DMO}$	& -0.052 \\ 
		$T_{\rm FP}$	& -0.20	& $T_{\rm DMO}$	& -0.14	 \\ 
		\hline
	\end{tabular*}
	\caption{
		Spearman correlation values  and $p$-value for shape parameters and the galaxy formation efficiency ($m_*/M_{200}$) for all haloes with  mass $>10^{11} M_\odot$, 
		corresponding to Figure~\ref{fig:galformcorr}.
	}
	\label{table:galformcorr}
\end{table}

To understand the relation between halo shape and a galaxy's stellar mass shown in Figure~\ref{fig:shape_mstar}, we examine the correlation between the inner halo shape and the galaxy formation efficiency.
Figure~\ref{fig:galformcorr} shows the inner halo shape measured at $r=0.15R_{200}$ as a function of $\log_{10} \left(m_*/M_{200}\right) $ for all haloes with $M_{200} > 10^{11} M_\odot$. Solid lines show the median parameters from Illustris  (red) with the 25th to 75th percentile as dashed lines, and the 2D histogram in the background gives the relative number density of haloes/galaxies in the considered parameter space. 
For comparison, we also plot the halo shapes from Illustris-Dark (black), assigning these haloes stellar masses based on their matched counterparts in Illustris-1.
To determine how strong the correlation is for each curve, we calculate the Spearman correlations and show them in Table \ref{table:galformcorr}.

In Illustris, we find that the axis ratios $q$ and $s$ vary substantially, and correlate positively with the galaxy formation efficiency.
Haloes with high galaxy formation efficiency are most spherical, 
with $q\approx 0.95$ and $s\approx 0.8$ when $m_*/M_{200}=0.1$.
However, such a correlation by itself is insufficient to show that rounder halo shapes is a direct result of higher galaxy formation efficiency.
It is also possible that galaxies with large $m_*/M_{200}$ only form in haloes which were originally already more spherical.
Our results using the halo shapes from matched haloes in Illustris-Dark show that this is not possible,
since the curves are flat, and the small Spearman correlation values indicate little correlation between the axis ratios and the DMO halo shapes.
Although we have included all haloes with resolved shapes here, we verified that this relation holds even when we examined haloes in smaller mass bins.

Even in haloes with the smallest galaxy formation efficiency,
(stellar-to-halo mass ratio of 0.01), 
we find that baryons still exert a noticeable impact in sphericalising the haloes.
Although our simulation box does not contain haloes with smaller stellar-to-halo mass ratios,
we expect the difference between the Illustris and Illustris-Dark shapes to shrink and become negligible for  haloes with mass $< 10^{11} M_\odot$.
Our results are similar to that of \cite{Butsky16v462}, extending the results to a larger sample of haloes and much larger halo masses.
\cite{Butsky16v462} examined zoomed-in haloes between $10^{10}$ and $10^{12} M_\odot$ and were able to resolve small galaxies with smaller masses and lower galaxy formation efficiency ($m_*/M_{200} < 0.01$) than in Illustris.  At these low efficiencies, they found that the impact of baryons is indeed minimal.
The convergence of our results despite the different hydrodynamic solvers 
and galaxy-physics implementations 
is a very good indication that the dependence of the halo shape on the galaxy formation efficiency is robust.

\section{Summary and Discussion}

In this paper, we have analysed the Illustris simulation suite to quantify the impact of galaxy formation on the shape of DM haloes. The Illustris suite includes a full hydrodynamical, galaxy-physics simulation (Illustris) and an equivalent dark matter-only (DMO) simulation (Illustris-Dark), each at three different resolutions. At the highest resolution ($2\times 1820^3$ elements in Illustris), we are able to study over 10,000 haloes with masses between $10^{11} $ and $ 3\times10^{14} M_\odot$. Instead of inferring and using a single value to characterize the shape of a halo, we have measured DM halo shapes in ellipsoidal shells at radii between $0.01 R_{200}$ and $R_{200}$. Our procedure utilized the unweighted shape tensor to measure the axis ratios $s \equiv c/a$ and $q \equiv b/a$, as well as the triaxiality $T \equiv (1 - q^2)/( 1- s^2)$ for each halo ($a>b>c$).  Our main results are summarized as follows:

\begin{enumerate}
	\item We have performed resolution tests to determine the convergence of the shape profiles $s(r)$ and $q(r)$, using the DMO runs,  which contain $1820^3$, $910^3$ and $455^3$ DM particles for the high, middle and low resolution run, respectively. We find that the shape profiles are converged only for  $r > r_{\rm conv} = 9 \epsilon$, where $\epsilon$ is the Plummer-equivalent softening length of DM particles in the simulations (Figure~\ref{fig:resolutiontest}). Our value of $r_{\rm conv}$ is larger than the value \cite{Power2003v338} determined for the convergence of halo mass profiles. For $10^{11} \msun$ haloes in the high resolution run, this corresponds to a radius close to 10 per cent of the virial radius, comfortably smaller than our reference inner-halo radius: $0.15R_{200}$.\\
    
    \item We have compared the halo shapes of the middle and low resolution runs of Illustris-Dark to their non-radiative counterpart (Illustris-NR, not available at the highest resolution), and find no differences (Figure~\ref{fig:NR}). Namely, we find that the evolution of gas elements through the moving-mesh hydrodynamics alone does not cause any changes in DM halo shapes in the absence of other galaxy formation physics such as radiative cooling and heating, star formation, and feedback.\\
	
	\item From our full galaxy-physics run, we find instead baryonic physics to have a significant impact on the halo shape throughout the halo, sphericalising haloes and causing them to become more oblate at a given radius (Figures~\ref{fig:shape_all} and \ref{fig:corr_dmfp}). This effect is strongest in the inner halo (defined here as $0.15R_{200}$), where the median axis ratios $s\equiv c/a$ and $q\equiv b/a$ in Illustris are increased by 0.2 points from their DMO values (Figure~\ref{fig:shape_m200}). The effects of baryons decrease away from the halo centre, hence the shape parameters at the virial radius are similar between Illustris and Illustris-Dark. These statements apply to the average galaxy or halo population, but some non-negligible scatter in the baryonic effects can still be appreciated (Figure~\ref{fig:corr_dmfp}).\\
    
    \item Baryons alter the orbital structure of haloes by increasing the DM velocity dispersions and decreasing the velocity anisotropies across all radii and masses, which means that orbits become more tangentially biased (Figure~\ref{fig:beta}). Unlike for halo shapes, where the Illustris and Illustris-Dark distributions of shape parameter values are well-separated across the galaxy population, there is substantial overlap between the velocity anisotropies of the two runs for haloes of similar mass.\\
	
	\item By focusing on MW-analogues of mass $M_{200} \approx 10^{12} M_\odot$ in Illustris, we find the DM halo shape parameters to read on average: $q_{\rm FP}  = 0.88 \pm 0.10$ and $s_{\rm FP}  = 0.70 \pm 0.11$ in the inner halo. This compares to $q_{\rm DMO}  = 0.67 \pm 0.14$ and $s_{\rm DMO}  = 0.52 \pm 0.10$ for MW-analogues in Illustris-Dark (Figure~\ref{fig:MWshape}). The resulting distribution of parameters in Illustris somewhat improves the agreement between numerical simulations and observational measurements of the MW halo shape.\\
	
	\item For comparison with observations which assume that the halo minor axis is perpendicular to the stellar disk, we derive in Illustris the axis ratios $q_{\rm fixed}$ and $s_{\rm fixed}$ with the halo minor axis constrained in the direction of the stellar spin. We find that misalignments between the stellar and DM halo minor axes result in a mean increase in the apparent value of $s$ by $\approx 0.11$, and the appearance of haloes with $s_{\rm fixed} > 1$ (Figure~\ref{fig:MWshape}).\\
		
	\item For MW-like haloes, we demonstrate that Illustris largely retains the correlations from Illustris-Dark between halo shape and halo formation time as well as between halo shape and velocity anisotropy. In Illustris, $q$ and $s$ correlate with formation time and concentration, and anti-correlate with velocity anisotropy. Interestingly, halo shape correlates with DM halo concentration somewhat more strongly in Illustris, whereas such relation is essentially absent in Illustris-Dark (Figure~\ref{fig:halocorr}).\\
	
    \item Finally, at a fixed fraction of the virial radius, the axis ratios $q$ and $s$ of the inner halo decrease monotonically as a function of halo mass in both Illustris and Illustris-Dark, similar to the results from previous $N$-body simulations. Conversely, the shape parameters become strongly non-monotonic with stellar mass, attaining their maximum values for haloes with $m_* = 10^{10.5-11} M_\odot$. For our galaxy formation implementation, these haloes are the most spherical and oblate. The dependence on the stellar mass is best explained by the galaxy formation efficiency, which we found to correlate strongly with the inner halo shape parameters (Figure~\ref{fig:galformcorr}).
    
\end{enumerate}

Our results are qualitatively consistent with those from previous hydrodynamic studies by e.g. \cite{Abadi10v407} and \cite{Butsky16v462} who note the roughly radius-independent shapes of haloes but for smaller masses ($M_{200} <10^{12} M_\odot$.)
These simulations lack larger haloes due to their small simulation volumes.
In particular, \cite{Abadi10v407} found in their zoomed-in hydrodynamic re-simulations of 13 MW-sized haloes that $s$ was roughly constant in halocentric distance, with a value of $\left< s\right> \approx 0.85$.
The authors note that this value is an upper bound for $s$ because their simulations neglect feedback, leading to unrealistically large galaxies. 
With the more realistic galaxy formation implementation in Illustris, our result for halo of the same mass $\left(\left< s\right> \sim 0.7\right)$ is in line with their expectations.

The increase in both sphericity and oblateness of DM haloes in full-physics simulations can be explained by the condensation of baryons into the centre of the haloes. 
The central baryonic mass scatters DM particles that approach the halo centre, modifying their orbits into rounder passages.
\cite{Debattista08v681} found using controlled numerical experiments that growing a central component in a halo can destroy box orbits, turning them into more circular tube orbits, consistent with our findings on the DM velocity anisotropy .
A similar conclusion was found by \cite{Barnes96v471}, who used idealized simulations of galaxy
interactions to study the orbital structure of merger remnants with and without 
gas.

While the baryons in e.g. Illustris do seem to reduce the tension between numerical simulations and observations of the MW stellar streams, possible inconsistencies with certain observations many continue to exist. In particular, the results of \cite{Bovy16v833} suggest a sphericity value that is improbable (at the 1$\sigma$ level) for Illustris galaxies after allowing for misalignments between the stellar and DM halo shapes, and assuming a good match between the simulated stellar and halo masses with those of the Galaxy. Such tensions could prove to be invaluable in evaluating if a modification to the $\Lambda$CDM framework is required. Such modifications can include 1) warm dark matter (WDM) models \citep[e.g.][]{Lovell12v420}, 2) self-interacting dark matter (SIDM) models \citep[e.g.][]{Vogelsberger16v460}, and 3) `fuzzy' cold dark matter models where dark matter is comprised of ultra-light ($m_{\rm DM} \sim 10^{-22}$ eV) scalar-field particles \citep[e.g.][]{Hu00v85,Marsh13v437}, all of which have been introduced to explain cored density profiles of the MW dwarf satellites, but could also lead to additional sphericalisation within the inner haloes of more massive systems as well. Finally, further quantitative comparisons between models and observationally-derived constraints can shed light on aspects of galaxy-physics models, as different subgrid-physics implementation may give rise to quantitatively (or even qualitatively) different effects of baryons on the phase-space properties of dark matter.

\section*{Acknowledgements}
We thank Frank van den Bosch for constructive comments on the paper. The simulations analyzed in this paper were run on the Harvard Odyssey and CfA/ITC clusters, the Ranger and Stampede supercomputers at the Texas Advanced Computing Center as part of XSEDE, the Kraken supercomputer at Oak Ridge National Laboratory as part of XSEDE, the CURIE supercomputer at CEA/France as part of PRACE project RA0844, and the SuperMUC computer at the Leibniz Computing Centre, Germany, as part of project pr85je. K.T.E.C. acknowledges support from the Singapore National Science Scholarship.


\appendix
\section{Convergence tests and modifications to the shape algorithm}
\begin{figure}
	\centering
	\includegraphics[width=0.47\textwidth]{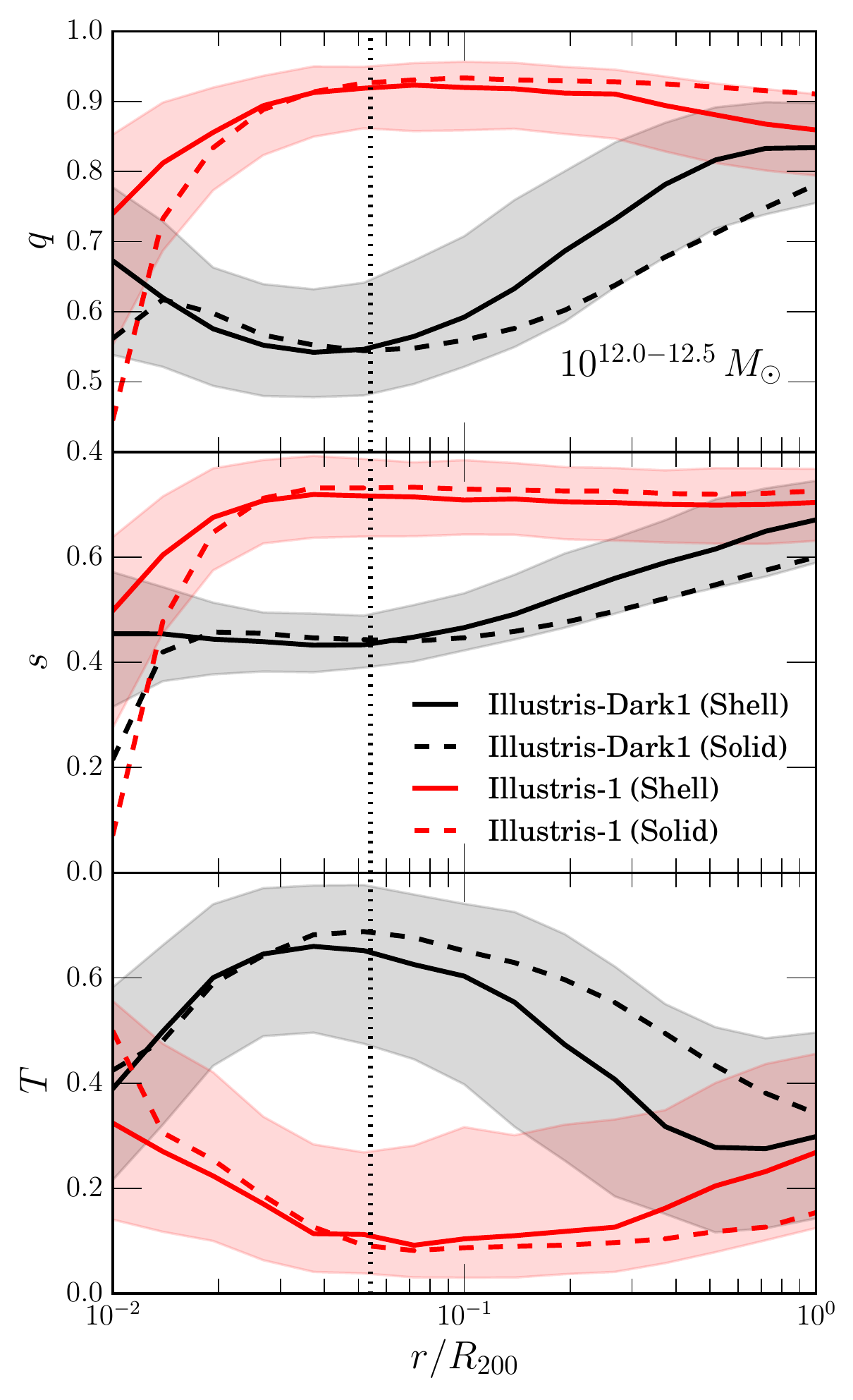}
	\caption{
		Comparison of inferred halo shape profiles for haloes of mass $10^{12-12.5} M_\odot$ using 1) the local halo shape of ellipsoidal shells (solid line, as adopted in the rest of this work), and 2) the reduced inertia tensor using all particles enclosed within an ellipsoidal volume (dashed line). Results for Illustris and Illustris-Dark are shown in red and black respectively. The difference between the inferred shapes using both methods is minor at small radii, but increases towards the virial radius.
	}
	\label{fig:shell_vs_volume}
\end{figure}

\begin{figure}
	\centering
	\includegraphics[width=0.47\textwidth]{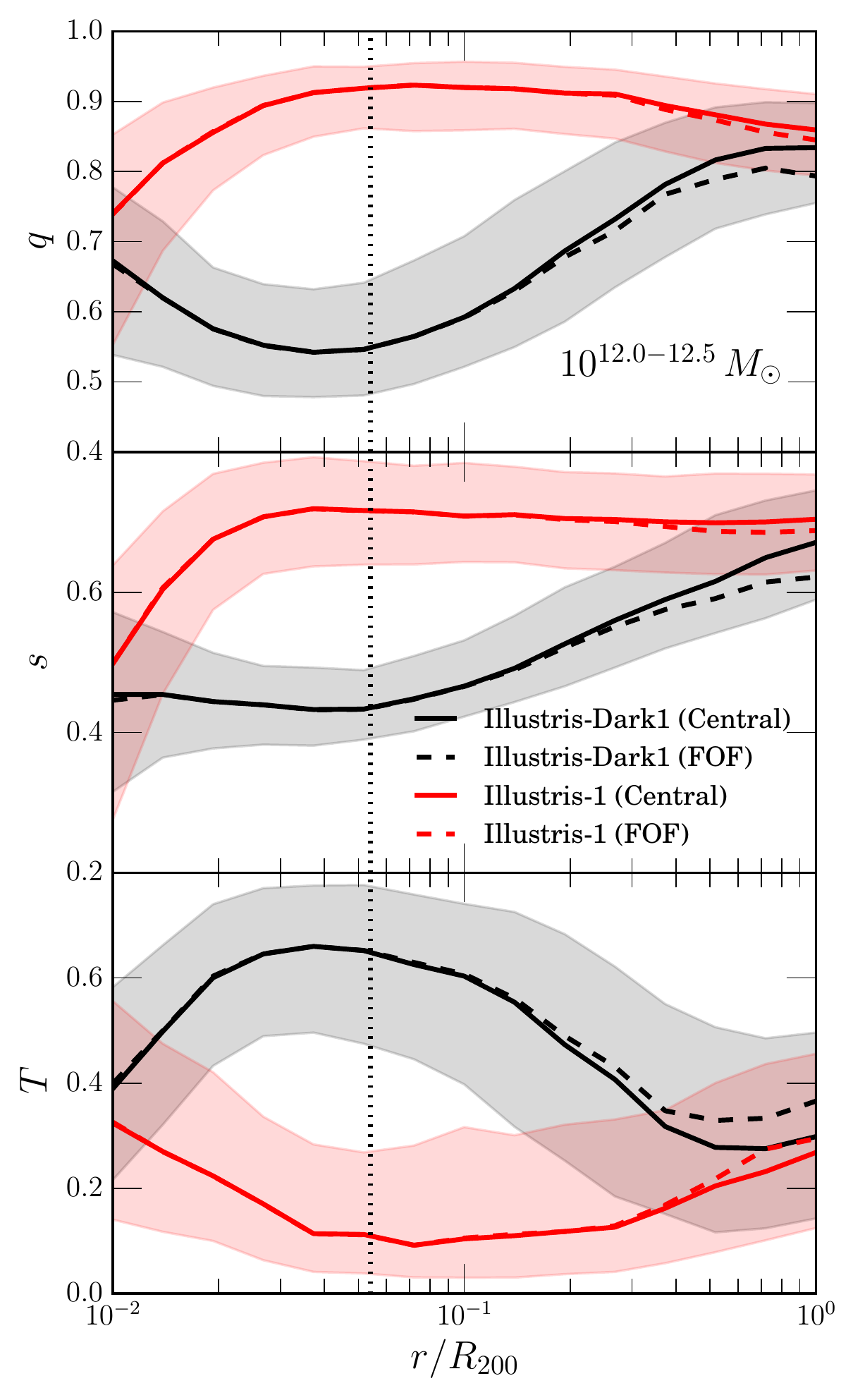}:
	\caption{
		Effect of substructure on halo shape profiles showing the shape parameters $s$ (top), $q$ (middle) and $T$ (bottom) as a function of halocentric distance for haloes of mass $10^{12-12.5} M_\odot$. Results for Illustris and Illustris-Dark are shown in red and black respectively. Solid lines indicate shapes calculated using only the central subhalo while dashed lines indicate shapes calculated using all particles identified to be part of the FOF group (thus including substructure.) Substructure have a noticeable effect only near the virial radius, decreasing sphericity and increasing the prolateness of the haloes.
	}
	\label{fig:substructure}
\end{figure}

\begin{figure*}
	\centering
	\includegraphics[width=0.95\textwidth]{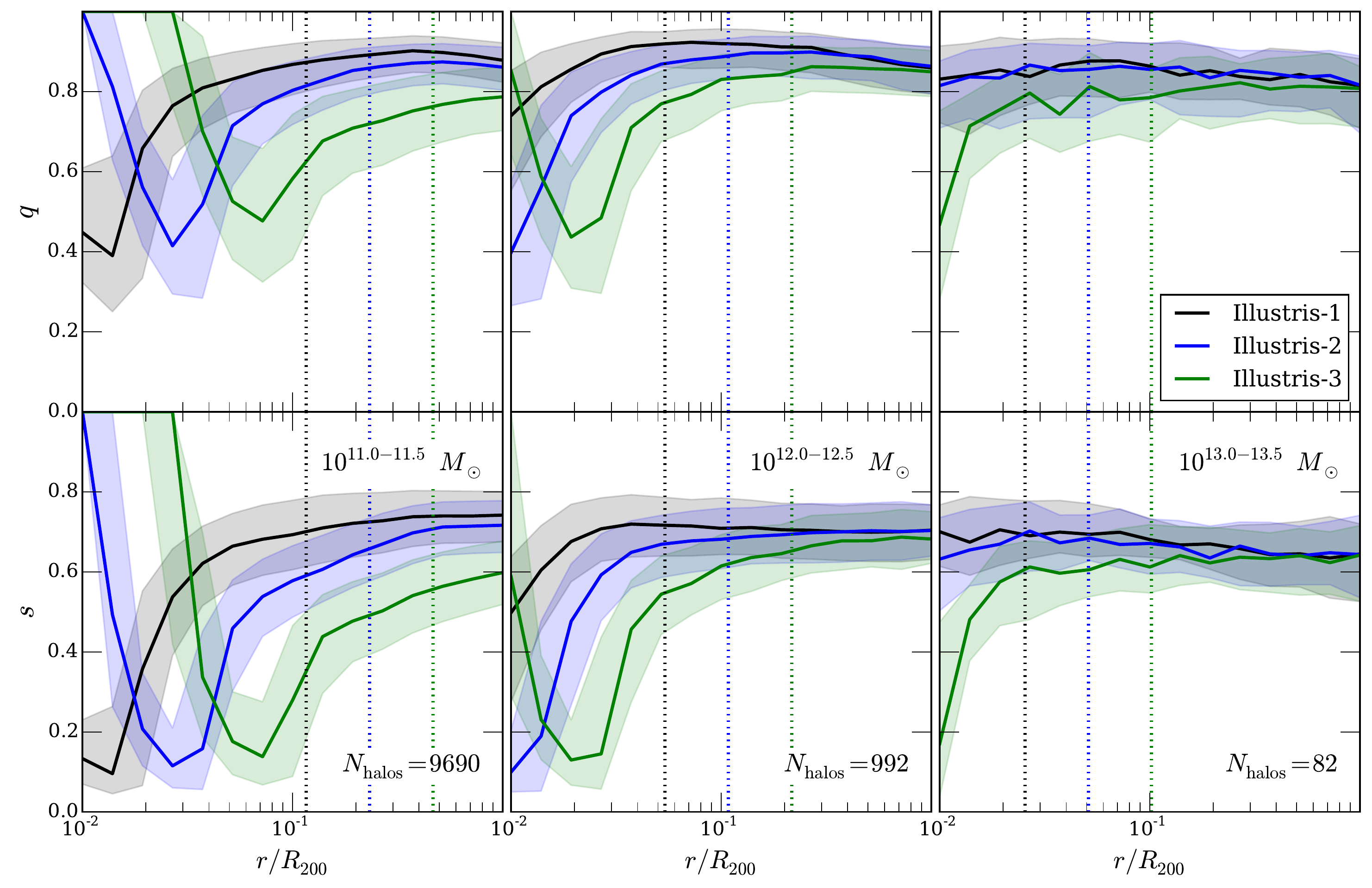}
	\caption{
		Dependence of shape profiles with resolution in the full-physics Illustris runs. Labels are similar to those shown for the DMO runs in Figure~\ref{fig:resolutiontest}, with Illustris-1 being the highest resolution run and Illustris-3 being the lowest resolutio run.
		As resolution affects stellar-to-halo mass relation, the shape profiles are less well converged than the DMO Illustris-Dark results of Figure~\ref{fig:resolutiontest}. 
		The number of haloes identified from each mass bin in Illustris-1 is also shown in the bottom row.
        }
	\label{fig:resolutiontestFP}
\end{figure*}

\subsection{Difference between ellipsoidal shells and volumes in shape determination}

Section \ref{sec:iterative} discussed various ways in which the shape tensor (Equation \ref{eqn:shapetensor}) can be utilized. For example, $S_{ij}$ can weighted or unweighted, and the DM particles can chosen either from an enclosed ellipsoidal volumes or from ellipsoidal shells. For the majority of this work, the halo shape is synonymous with the \emph{local} shape, which is determined using an unweighted shape tensor with thin ellipsoidal shells. On other hand, when an enclosed ellipsoidal volume is desired, it is preferable to use the reduced shape tensor (with weights $w_k = r_{\rm ell}^2$), which reduces the contribution of particles at large radii to the shape tensor. Since both are iterative methods which have been employed in literature, we examine here the difference between both methods, concentrating on $10^{12-12.5} \msun$ haloes in both Illustris and Illustris-Dark.

Figure~\ref{fig:shell_vs_volume} shows the inferred halo shapes for the two methods, with solid lines representing the local shapes and dashed lines representing shapes calculated using the reduced inertial tensor. In Illustris-Dark, using the reduced inertial tensor biases the inferred values of $q$ and $s$ towards larger values. Conversely in Illustris, the inferred values of $q$ and $s$ are biased towards smaller values. The difference between the two methods is negligible in the inner halo and increases towards the virial radius. These results can be traced to the definition of the reduced inertial tensor, which weights the inner particles more strongly, thus biasing the inferred shapes towards that of the inner halo. As a result, the profiles are 1) smoothed out compared to the local shape, and 2) any changes in shape also lag behind the latter. These conclusions are similar to those found by \cite{Zemp11v197}.

\subsection{Effect of satellites in shape determination}

While we default to calculating shapes using only the central subhalo in each halo, we briefly examine the impact of including substructure by using all particles belonging to the halo (or FOF group). Figure~\ref{fig:substructure} shows the effect of substructure on halo shape profiles for $10^{12} M_\odot$ haloes in Illustris and Illustris-Dark.

We find that including subhaloes causes the inferred halo shape to be less spherical (lower $q$ and $s$) and more prolate (higher $T$). This effect is only noticeable near the virial radius and decreases toward the halo centre. Subhaloes, being gravitationally bound clumps of dark matter and baryons, distort the shape tensor and biases the inferred parameters to being less spherical.
The increasing effect of subhaloes with radius reflects the increasing subhalo mass fraction with radius in haloes \citep{Aquarius}. Similarly, the effect of subhaloes is also smaller in Illustris compared to Illustris-Dark due to the decreased abundance of subhaloes in Illustris \citep{Chua2017}.

\subsection{Effect of resolution on halo shapes in the presence of baryons}
\label{sec:IllustrisReso}

Section \ref{sec:resolution} discussed the convergence of shape profiles for the three resolution runs of the DMO Illustris-Dark. As mentioned, the reason for doing so was to ignore shape changes due to the dependence of baryonic effects on resolution. In Illustris, it has been observed that both the star formation rate and the  stellar mass of galaxies decrease with decreasing resolution \citep{Vogelsberger13}.

To illustrate the resolution dependence of baryonic effects on halo shapes, we show in Figure~\ref{fig:resolutiontestFP} the median halo shape profiles for Illustris haloes in the three resolution runs. Compared to the DMO results (as shown in Figure~\ref{fig:resolutiontest}), the Illustris profiles are noticeably less converged across resolutions. The lowest resolution run Illustris-3 is not well converged with both higher resolution runs, even for $10^{13-13.5} \msun$ haloes. For small $10^{11-11.5} \sun$ haloes, Illustris-2 demonstrates the same issue as Illustris-3: the median shape profiles are also in disagreement with Illustris-1. The agreement between Illustris-2 and Illustris-1 is improved for more massive systems, although the convergence remains poorer than that observed in the DMO runs. As such, we find that shape convergence in Illustris depends strongly not only on resolution, as observed in the DMO runs, but also on halo mass. The poor convergences can be traced to the under-prediction of galaxy stellar mass and galaxy formation efficiency at lower resolutions \citep{Vogelsberger13}, especially for lower mass haloes, thus leading to the formation of less spherical haloes (as discussed in Section \ref{sec:galform}). Evidently, the convergence of halo shapes with resolution is more complicated when baryonic physics is introduced, and is likely dependent on the specific galaxy formation implementation adopted in the hydrodynamic simulation. For this reason, we have chosen to focus on the DMO runs for our resolution tests in Section \ref{sec:resolution}.

{\bibliographystyle{mnras}
\bibliography{references}
}

\end{document}